\documentclass[useAMS,usenatbib]{mn2e}
\usepackage{graphicx}
\usepackage{amssymb}
\usepackage{color}

\title[Spatial distributions of core-collapse supernovae in infrared-bright galaxies]{Spatial distributions of core-collapse supernovae in infrared-bright galaxies}
\author[Kangas et al.]{T. Kangas$^{1,2}$\thanks{E-mail:
tjakan@utu.fi or tjakan@not.iac.es}, S. Mattila$^{3}$, E. Kankare$^{3}$, J. K. Kotilainen$^{3}$, P. V\"{a}is\"{a}nen$^{4}$, R. Greimel$^{5}$ \and and A. Takalo$^{1}$  \\
$^{1}$Tuorla Observatory, Department of Physics and Astronomy, University of Turku, V\"{a}is\"{a}l\"{a}ntie 20, FI-21500 Piikki\"{o}, Finland\\
$^{2}$Nordic Optical Telescope, Apartado 474, E-38700 Santa Cruz de La Palma, Spain\\
$^{3}$Finnish Centre for Astronomy with ESO (FINCA), University of Turku, V\"{a}is\"{a}l\"{a}ntie 20, FI-21500 Piikki\"{o}, Finland\\
$^{4}$South African Astronomical Observatory, PO Box 9, 7935 Observatory, Cape Town, South Africa \\
$^{5}$Institut f\"ur Physik, Universit\"at Graz, Universit\"atsplatz 5/II, A-8010 Graz, Austria \\}

\begin{document}

\date{2013}

\pagerange{\pageref{firstpage}--\pageref{lastpage}} \pubyear{2013}

\maketitle

\label{firstpage}

\begin{abstract}
We have measured the correlation between the locations of core-collapse supernovae (CCSNe) and host galaxy light in the H$\alpha$ line, near ultraviolet (NUV), \emph{R}-band and \emph{Ks}-band to constrain the progenitors of CCSNe using pixel statistics. Our sample consists of 86 CCSNe in 57 infrared (IR) -bright galaxies, of which many are starbursts and ten are luminous infrared galaxies (LIRGs). We also analyse the radial distribution of CCSNe in these galaxies, and determine power-law and exponential fits to CCSN surface density profiles. To probe differences between the SN population of these galaxies and normal spiral galaxies, our results were compared to previous similar studies with samples dominated by normal spiral galaxies where possible. We obtained a normalised scale length of $0.23^{+0.03}_{-0.02}$ $R_{25}$ for the surface density of CCSNe in IR-bright galaxies; less than that derived for CCSNe in a galaxy sample dominated by normal spiral galaxies (0.29 $\pm$ 0.01). This reflects a more centrally concentrated population of massive stars in IR-bright galaxies. Furthermore, this centralisation is dominated by a central excess of type Ibc/IIb SNe. This may be due to a top-heavy initial mass function and/or an enhanced close binary fraction in regions of enhanced star formation. Type Ic SNe are most strongly correlated with H$\alpha$ light and NUV-bright regions, reflecting the shortest lifetime and thus highest mass for type Ic progenitors. Previous studies with samples dominated by normal spiral galaxies have indicated a lower Ibc--H$\alpha$ correlation than our results do, which may be due to the central excess of type Ibc/IIb SNe in our sample. The difference between types II and Ib is insignificant, suggesting that progenitor mass is not the dominant factor in determining if a SN is of type Ib or II. Similar differences in correlation can be seen in the \emph{Ks}-band (which in these galaxies is dominated by red supergiants and thus also traces recent star formation), with type Ibc/IIb SNe tracing the \emph{Ks}-band light better than type II in our sample.

\end{abstract}


\section{Introduction}

Despite decades of ever-intensifying research on supernovae (SNe), our understanding of their exact origins still remains incomplete. The connection between the observed properties of CCSNe and the properties of their progenitors is one of the questions of interest - in particular the differences between the progenitors of different SN types; whether the stellar mass is the dominant parameter or whether other factors such as binarity, rotation or metallicity of the progenitor also contribute significantly (e.g. Smartt 2009; Smith et al. 2011; Eldridge et al. 2013 and references therein).

Supernovae are classified into two main types: type I, with no hydrogen lines in their spectra; and type II, whose spectra show these lines (e.g. Filippenko 1997).  SNe of types Ib, Ic (together referred to as type Ibc) and II, i.e. core-collapse supernovae (CCSNe), are thought to have their origin in massive stars ($\ge 8 M_{\odot}$, \citet{b12}) that explode after exhausting their nuclear fuel. The progenitors of type Ibc and IIb SNe are thought to be massive stars that have lost their outer envelopes (H envelope for Ib and both H and He for Ic) by stellar winds or through mass transfer to a binary companion (see e.g. Eldridge et al. 2013). The progenitors of other type II SNe, on the other hand, are believed to be less massive supergiant stars that have retained their envelopes. However, the nature of type IIn SN progenitors is unclear. At least part of them are believed to be extremely massive stars (e.g. Smith et al. 2011), but \citet{b21} instead found evidence that, on average, type IIn SNe originate from the lower mass end of the CCSN progenitors.

Attempts to directly observe CCSN progenitors in pre-explosion archive images of host galaxies have yielded direct detections of several SN II progenitors (e.g. Smartt et al. 2004; Li et al. 2006; Mattila et al. 2008; Van Dyk et al. 2011; Maund et al. 2011), but the need to be able to resolve individual stars in other galaxies typically requires a space telescope or adaptive optics. Another way to study the progenitors is to observe the environments of the SNe. Statistical analysis of galaxy light close to the locations of historical SNe can yield constraints on the nature of their progenitors. H$\alpha$ line emission is important in this context, since it traces regions of ionized hydrogen; and only massive young stars (that also give rise to CCSNe) contribute significantly to the ionizing radiation \citep{b11}.

\citet{b3} found an excess of type II SNe in regions of low star-forming activity in contrast to SNe of type Ibc by studying a sample of 63 SNe (of which, however, 25 were either of type Ia or had no designated type). This analysis was repeated for a total of 260 SNe in \citet{b20} and \citet{b21} (hereafter referred to as A08 and A12, respectively) with similar results. These results implied a mass sequence for CCSN progenitors running, in ascending order, from type II through Ib to Ic. \citet{b36} also argue for a higher mass for type Ibc progenitors than type II using colours of SN environments. In addition statistical studies of the radial distribution of CCSNe have indicated that type Ibc SNe are more centrally concentrated than type II SNe (e.g. Petrosian et al. 2005, Prieto, Stanek \& Beacom 2008, Anderson \& James 2009, Hakobyan et al. 2009). In particular, this also appeared to be the case for disturbed galaxies, with presumably higher star formation rates than in normal spiral galaxies (Habergham, Anderson \& James 2010; Habergham, James \& Anderson (2012). 

Anderson, Habergham \& James (2011) studied a single interacting luminous infared galaxy (LIRG; defined as a galaxy with $L_{IR} \geq 10^{11} L_{\odot}$) system, Arp 299 (NGC 3690 and IC 694), which has hosted 7 optically observed SNe. They found evidence that its population of SNe may be different from that of normal spiral galaxies -- with an abnormal number of stripped-envelope SNe abnormally close to the cores of the system -- although the study has to deal with small-number statistics. One possible interpretation is that, because of the young age of the most recent starburst episode in the system, insufficient time has elapsed for the observed SN rates to reflect the star formation rates for the full range of stellar masses capable of producing CCSNe $\gtrsim 8 M_{\odot}$. This lends support to type Ibc and IIb SNe arising from progenitors of shorter lifetime than other type II. On the other hand, the centralisation of type Ib/IIb SNe compared to other type II in Arp 299 -- and the central excess of type Ibc SNe found in other studies --implies that in the central region with enhanced star formation, the initial mass function (IMF) may favor high mass stars. According to \citet{b17}, the central excess found in normal spiral galaxies can alternatively be explained by higher metallicities of the progenitors of type Ibc SNe. Habergham et al. (2010, 2012) studied the central excess of type Ibc SNe in disturbed galaxies, considering also metallicity as a possible explanation, but found this possibility unlikely. 

The studies of e.g. \citet{b3} (as well as A08 and A12) and \citet{b17} mentioned above concern SNe in spiral galaxy hosts in general. In light of the recent results of \citet{b15}, we study CCSNe in infrared (IR) -bright galaxies in a similar way in this paper. Our goal is to find out whether the supernova and thus progenitor star population in these galaxies really differs from normal spiral galaxies. We apply the methods used by \citet{b3} (and in subsequent papers), extending them to the \emph{R}- and \emph{Ks}-bands and to the near-UV. We also apply the method of \citet{b17}. The results are then compared to the earlier studies. The paper is arranged as follows: in Section 2 we describe our sample and observations; Section 3 presents our data reduction methods; Sections 4 and 5 cover the statistical tools at our disposal and an overview of our results based on each method, respectively; in Section 6, we discuss our results; and finally, in Section 7, we present our conclusions.

\section{Sample and Observations}

A sample of IR-bright galaxies was selected from among those in the IRAS Revised Bright Galaxy Catalog \citep{b22} based on their far-IR luminosities ($L_{FIR}$, in the wavelength range of 40 -- 400 $\mu$m) and distance following the criteria below. Far-infrared luminosity was used as a criterion since it is a good tracer of star formation in actively star-forming galaxies. Luminosities and distances were taken from \citet{b22}. The value for the Hubble constant used in this catalog was 75 km s$^{-1}$ Mpc$^{-1}$. \\ \\
(1) $d < 75$ Mpc, \\
(2) $L_{FIR} > 1.6 \times 10^{10} L_{\odot}$ (log $L_{FIR} > 10.2$). \\
(3) Type 1 Seyfert galaxies listed in \citet{b8} were excluded to avoid contamination by an active galactic nucleus (AGN). \\
(4) Observable from La Palma with a reasonable airmass, i.e. Decl. $\gtrsim -35^{\circ}$. \\

\citet{b31} argue that after $cz > 6000$ km s$^{-1}$ one will begin to miss SNe with low projected distance from the galaxy core. With our distance limit, this value is not exceeded by any host galaxy. The limit in $L_{FIR}$ is roughly half of that of the `prototypical' starburst galaxy, M 82, in order to include a larger number of galaxies that still have high star formation rates (SFRs). It corresponds to a star formation rate of $\sim 2.7 M_{\odot}$ yr$^{-1}$ according to eq. 4 in \citet{b11} and a CCSN rate of $\sim 0.04$ yr$^{-1}$ according to eq. 1 in \citet{a01}. The median distance in the sample is $\sim 36$ Mpc and the median $L_{FIR}$ is 3.7 $\times 10^{10} L_{\odot}$. A comparison between our galaxy sample and that of A08/A12 is presented in Fig. 1, to show the different distance and luminosity ranges covered. The majority of the A08/A12 sample is well below our luminosity limit.

\begin{figure}
\centering
\begin{minipage}{83mm}
\includegraphics[width=\textwidth]{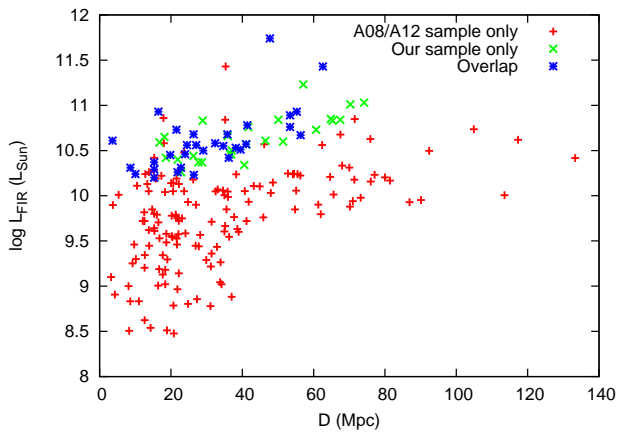}
\caption{A comparison between the luminosities and distance ranges of our host galaxy sample and that of A08 and A12.}
\end{minipage}
\end{figure}

Any SNe discovered before 1990 were excluded from the sample. These SNe mostly had no (sub)type classification, and their exact coordinates were considered possibly inaccurate because they were often discovered using photographic plates. The coordinates of the SNe were obtained from the Asiago supernova catalog\footnote{http://heasarc.gsfc.nasa.gov/W3Browse/all/asiagosn.html}. All SNe of type Ia were also excluded, since they arise from a much older stellar population than CCSNe and are thus outside the scope of this paper. This resulted in a total sample of 57 galaxies -- 22 of which are brighter in the FIR than M82, the prototypical starburst galaxy, and ten of which are LIRGs. The galaxy sample was sorted into three classes by eye, based on the presence or lack of tidal features or companion galaxies within a few arcminutes: isolated (28 galaxies), ``disturbed'' (19 galaxies) and merger/close pair (10 galaxies). The SN sample consists of 86 SNe, of which 35 were of type Ib, Ic or IIb (\emph{stripped-envelope SNe (SE-SNe)}); 47 were of other type II (II-P, II-L, IIn and those only classified `II'); and four were of undetermined type. For a complete list of CCSN host galaxies included in the study, see Table 1.

\begin{table*}
\centering
\begin{minipage}{177mm}
\caption{A list of our sample galaxies. `Telescope' given in the form optical/NIR, if both were not observed at the NOT. Seeing FWHM is also given in the form optical/NIR. RA, Decl, distance, $L_{FIR}$ from \citet{b22}; $R_{25}$ and $i$ from HyperLeda or NED. The $R_{25}$, $i$ and CCSNe values for galaxy pairs are given for the respective components.}
\begin{tabular}{lccccccccc}
    \hline
	  Galaxy & RA        & Decl        & Telescope & FWHM &  D          & lg($L_{FIR}$) & $R_{25}$ & $i$ & CCSNe \\
         & (J2000) & (J2000) & & ('') & (Mpc)       & ($L_{\odot}$) & ('') & ($^{\circ}$) & \\
    \hline
    NGC 157 & 00 34 46.0 & -08 23 53 & NOT & 1.2/0.8 & 21.92 & 10.40 & 111 & 61.8 & 1\\
    UGC 556 & 00 54 49.9 & +29 14 42 & NOT & 0.9/0.7 & 60.71 & 10.73 & 28 & 60.9 & 1\\
    NGC 317B & 00 57 40.9 & +43 47 37 & NOT & 1.1/0.6 & 70.3 & 11.01 & 29 & 65.5 & 1\\
    NGC 772 & 01 59 19.5 & +19 00 23 & NOT & 1.7/0.9 & 28.71 & 10.37 & 137 & 59.6 & 2\\
    NGC 838 & 02 09 38.8 & -10 08 46 & NOT & 1.0/0.9 & 50.13 & 10.84 & 40 & 49.8 & 1\\
    NGC 922 & 02 25 04.2 & -24 47 21 & NOT & 0.9/1.0 & 40.57 & 10.34 & 63 & 33.4 & 2\\
    NGC 1084 & 02 45 59.7 & -07 34 41 & NOT & 0.8/0.8 & 18.61 & 10.42 & 102 & 49.9 & 3\footnote{SN 2012ec located in NGC 1084 was still visible when the galaxy was observed, and thus could not be included in this study.}\\
    NGC 1097 & 02 46 18.7 & -30 16 29 & NOT & 1.2/0.7 & 16.8 & 10.59 & 314 & 55.0 & 3 \\
    NGC 1614 & 04 34 00.1 & -08 34 46 & NOT & 0.9/0.7 & 62.61 & 11.43 & 38 & 41.8 & 1\\
    NGC 1961 & 05 42 03.9 & +69 22 33 & NOT & 0.9/0.8 & 55.27 & 10.93 & 131 & 47.0 & 1\\
    NGC 2146 & 06 18 39.8 & +78 21 25 & INT/WHT & 1.3/0.8 & 16.47 & 10.93 & 161 & 37.4 & 1\\
    ESO 492-G02 & 07 11 40.7 & -26 42 19 & NOT & 1.4/0.6 & 36.22 & 10.42 & 63 & 48.8 & 1\\
    UGC 3829 & 07 23 43.1 & +33 26 32 & NOT & 1.1/0.8 & 56.37 & 10.67 & 31 & 40.5 & 1 \\
    NGC 2276 & 07 27 17.8 & +85 45 16 & NOT & 0.9/1.0 & 35.83 & 10.68 & 67 & 39.8 & 2\\
    NGC 2415 & 07 36 56.0 & +35 14 31 & NOT & 0.9/0.9 & 53.41 & 10.76 & 25 & 0.0 & 2\\
    NGC 2782 & 09 14 05.7 & +40 06 47 & NOT & 1.1/0.5 & 39.51 & 10.51 & 97 & 45.2 & 1\\
    NGC 2993 & 09 45 47.3 & -14 21 00 & NOT & 0.9/0.9 & 34.69 & 10.55 & 40 & 35.8 & 1\\
    IC 2522 & 09 55 07.3 & -33 08 04 & NOT & 1.2/1.0 & 41.16 & 10.57 & 69 & 41.9 & 2\\
    M 82 & 09 55 53.1 & +69 40 41 & KP & 1.0/1.2 & 3.63 & 10.61 & 329 & 76.9 & 2\\
    NGC 3095 & 10 00 06.2 & -31 33 15 & NOT & 1.1/0.9 & 36.86 & 10.46 & 109 & 62.1 & 2\\
    NGC 3094 & 10 01 25.5 & +15 46 15 & NOT/WHT & 1.3/1.3 & 37.04 & 10.50 & 41 & 46.2 & 1\\
    NGC 3079 & 10 01 57.9 & +55 40 51 & NOT & 0.7/0.7 & 18.19 & 10.65 & 244 & 90.0 & 1\\
    NGC 3147 & 10 16 55.4 & +73 23 54 & NOT & 0.9/0.6 & 41.41 & 10.78 & 122 & 31.2 & 1\\
    NGC 3310 & 10 38 46.2 & +53 30 08 & NOT/WHT & 0.9/1.6 & 19.81 & 10.45 & 57 & 16.1 & 1\\
    NGC 3504 & 11 03 11.1 & +27 58 22 & NOT/WHT & 2.5/1.2 & 27.07 & 10.56 & 74 & 26.1 & 2\\
    M 66 & 11 20 14.9 & +12 59 30 & KP/NOT & 1.2/0.4 & 10.04 & 10.24 & 307 & 67.5 & 2\\
    NGC 3655 & 11 22 53.5 & +16 35 28 & NOT & 0.7/0.9 & 26.42 & 10.23 & 45 & 47.1 & 1 \\
    NGC 3672 & 11 25 01.6 & -09 47 34 & NOT & 1.5/0.7 & 27.70 & 10.37 & 87 & 56.2 & 1\\
    NGC 3683 & 11 27 30.6 & +56 52 44 & NOT & 0.8/0.8 & 29.06 & 10.50 & 52 & 68.8 & 1\\
    NGC 3690/IC 694 & 11 28 30.4 & +58 34 10 & NOT/WHT & 1.3/0.6 & 47.74 & 11.74 & 72/35 & 43.6/52.4 & 3/4\\
    NGC 4027 & 11 59 28.8 & -19 15 45 & NOT & 0.9/0.6 & 22.84 & 10.26 & 106 & 42.3 & 1\\
    NGC 4030 & 12 00 23.8 & -01 06 03 & NOT & 0.9/0.6 & 24.50 & 10.56 & 114 & 47.1 & 1\\
    NGC 4038/9 & 12 01 55.1 & -18 52 43 & NOT/WHT & 1.7/1.5 & 21.54 & 10.73 & 161/93 & 51.9/71.2 & 1/0\\
    NGC 4041 & 12 02 12.2 & +62 08 14 & NOT & 1.0/0.7 & 22.78 & 10.31 & 77 & 22.0 & 1\\
    M 61 & 12 21 55.4 & +04 28 24 & NOT & 1.8/0.7 & 15.29 & 10.37 & 208 & 18.1 & 3\\
    NGC 4433 & 12 27 37.5 & -08 16 35 & NOT & 0.7/0.6 & 41.68 & 10.76 & 60 & 79.6 & 1\\
    NGC 4527 & 12 34 09.9 & +02 39 04 & NOT & 1.8/0.6 & 15.29 & 10.29 & 189 & 81.2 & 1\\
    NGC 4567/8 & 12 36 33.7 & +11 14 32 & NOT & 1.5/0.7 & 15.29 & 10.20 & 83/128 & 39.4/67.5 & 0/2\\
    NGC 5020 & 13 12 39.1 & +12 36 07 & NOT & 0.9/0.9 & 51.45 & 10.60 & 81 & 27.0 & 1\\
    NGC 5078 & 13 19 50.0 & -27 24 37 & NOT & 1.3/0.9 & 26.32 & 10.44 & 77 & 90.0 & 1\\
    M 51 & 13 29 52.7 & +47 11 43 & KP/NOT & 1.3/0.5 & 8.63 & 10.31 & 414 & 32.6 & 3\\
    UGC 8739 & 13 49 15.0 & +35 15 17 & NOT & 0.9/0.6 & 74.18 & 11.03 & 57 & 87.5 & 1\\
    NGC 5371 & 13 55 39.1 & +40 27 34 & NOT & 1.2/0.6 & 41.06 & 10.57 & 119 & 54.0 & 1\\
    NGC 5394/5 & 13 58 36.0 & +37 25 58 & NOT & 0.7/0.6 & 53.40 & 10.89 & 52/75 & 70.8/66.1 & 0/1\\
    NGC 5433 & 14 02 37.4 & +32 30 27 & NOT & 1.2/0.7 & 65.30 & 10.83 & 45 & 90.0 & 1\\
    NGC 5597 & 14 24 26.5 & -16 45 43 & NOT/WHT & 0.6/0.9 & 36.68 & 10.46 & 60 & 40.0 & 1\\
    NGC 5775 & 14 53 58.0 & +03 32 32 & NOT & 1.0/0.6 & 26.34 & 10.68 & 111 & 83.2 & 1\\
    NGC 6000 & 15 49 49.4 & -29 23 11 & NOT/WHT & 5.9/1.3 & 28.90& 10.83 & 57 & 30.7 & 2\\
    NGC 6574 & 18 11 51.0 & +14 58 53 & NOT & 1.4/0.6 & 35.94 & 10.66 & 48 & 41.5 & 1\\
    NGC 6643 & 18 19 47.4 & +74 34 09 & NOT & 1.0/0.6 & 21.85 & 10.26 & 99 & 62.7 & 2\\
    NGC 6745 & 19 01 41.5 & +40 44 46 & NOT & 1.3/0.9 & 64.81 & 10.85 & 29 & 63.9 & 1\\
    NGC 6951 & 20 37 13.9 & +66 06 20 & NOT & 0.8/0.9 & 23.99 & 10.46 & 95 & 50.8 & 1\\
    NGC 7479 & 23 04 57.7 & +12 19 29 & NOT/WHT & 1.3/1.0 & 32.36 & 10.58 & 109 & 43.0 & 2\\
    NGC 7678 & 23 28 27.0 & +22 25 09 & NOT & 1.0/0.7 & 46.49 & 10.61 & 60 & 43.7 & 2\\
    NGC 7714 & 23 36 14.2 & +02 09 17 & NOT/WHT & 1.0/1.4 & 38.16 & 10.53 & 66 & 45.1 & 2 \\
    NGC 7753 & 23 47 01.3 & +29 28 11 & NOT & 1.4/0.6 & 67.46 & 10.84 & 60 & 82.1 & 1\\
    NGC 7771 & 23 51 24.7 & +20 06 39 & NOT & 0.8/0.8 & 57.11 & 11.23 & 75 & 66.7 & 1\\
\hline
\end{tabular}
\end{minipage}
\end{table*}

These galaxies were observed with the 2.56 m Nordic Optical Telescope (NOT) on La Palma, in the Canary Islands, using the Andalucia Faint Object Spectrograph and Camera (ALFOSC) instrument over several observing runs between 2009 and 2013. ALFOSC has a 2048 $\times$ 2048 pixel detector, with a $0\arcsec.19$ pixel scale and a field of view (FOV) of $6.4\arcmin \times 6.4\arcmin$. Integration times for H$\alpha$ on- and off-filter images were between 300 and 600 seconds, depending on target brightness. H$\alpha$ on- and off- filters were selected based on the redshift of the galaxy (either at 661.0 nm, for redshifts under 0.0103, or 665.3 nm for the remaining galaxies) and usually the on-filter includes the [N II] line as well. \emph{R}-band images for the sample were obtained using an exposure time of 100 s.

\emph{Ks}-band images for a majority of the sample were obtained using the Nordic Optical Telescope near-infrared Camera and spectrograph (NOTCam), which has a 1024 $\times$ 1024 pixel array, pixel scale of $0\arcsec.234$ and a FOV of $4\arcmin \times 4\arcmin$ in wide-field imaging mode. The images were taken with a nine-point dither pattern on-source and one minute of integration time per pointing. The rest of the \emph{Ks}-band images of the sample galaxies were obtained with the 4.2 m William Herschel Telescope (WHT), using the Isaac Newton Group Red Imaging Device (INGRID, Packham et al. 2003) or Long-slit Intermediate Resolution Infrared Spectrograph (LIRIS, Manchado et al. 2004) instruments, as a part of a near-IR SN search programme (see Mattila, Meikle \& Greimel 2004). INGRID has a 1024 $\times$ 1024 pixel array, with a $0\arcsec.24$ pixel scale and FOV of 4.2$\arcmin \times$ 4.2$\arcmin$; while LIRIS has a 1024 $\times$ 1024 pixel array and a scale of $0\arcsec.25$ yielding a FOV of 4$\arcmin.27 \times$ 4$\arcmin$.27. The WHT images were taken using a four-quarter dither pattern. The median seeing (FWHM) of the optical, \emph{Ks}-band and GALEX images in the sample are $1.0\arcsec$, $0.8\arcsec$ and $4.8\arcsec$, corresponding to roughly 170 pc, 140 pc and 830 pc at our median distance of 36 Mpc, respectively.

The H$\alpha$ and \emph{R}-band images for a few galaxies from \citet{b5} that matched our criteria were kindly shared by the authors: NGC 3034 (M 82), NGC 3627 (M 66), NGC 5194 (M 51) and NGC 4041. These galaxies had been observed with the Kitt Peak 2.1 m (KP) telescope. For the Kitt Peak H$\alpha$ images, a filter centered at 657.3 nm was used. In addition, the optical images of NGC 2146 were observed at the 2.5 m Isaac Newton Telescope (INT) with the Wide Field Camera (WFC), using an H$\alpha$ filter centered at 656.8 nm.

Near-ultraviolet (NUV) images were obtained from the GALEX space telescope data archive\footnote{http://galex.stsci.edu/GR6/default.aspx}. These images have a pixel scale of $1\arcsec.5$, roughly four times that of the 2 $\times$ 2 binned NOT/ALFOSC images used for this study, and a wavelength range of 177 -- 273 nm. GALEX data was not available for five galaxies in our sample: NGC 6000, NGC  6643, NGC 6745, UGC 3829 and ESO 492-G02.

\begin{figure*}
\centering
\begin{minipage}{170mm}
\includegraphics[width=17cm]{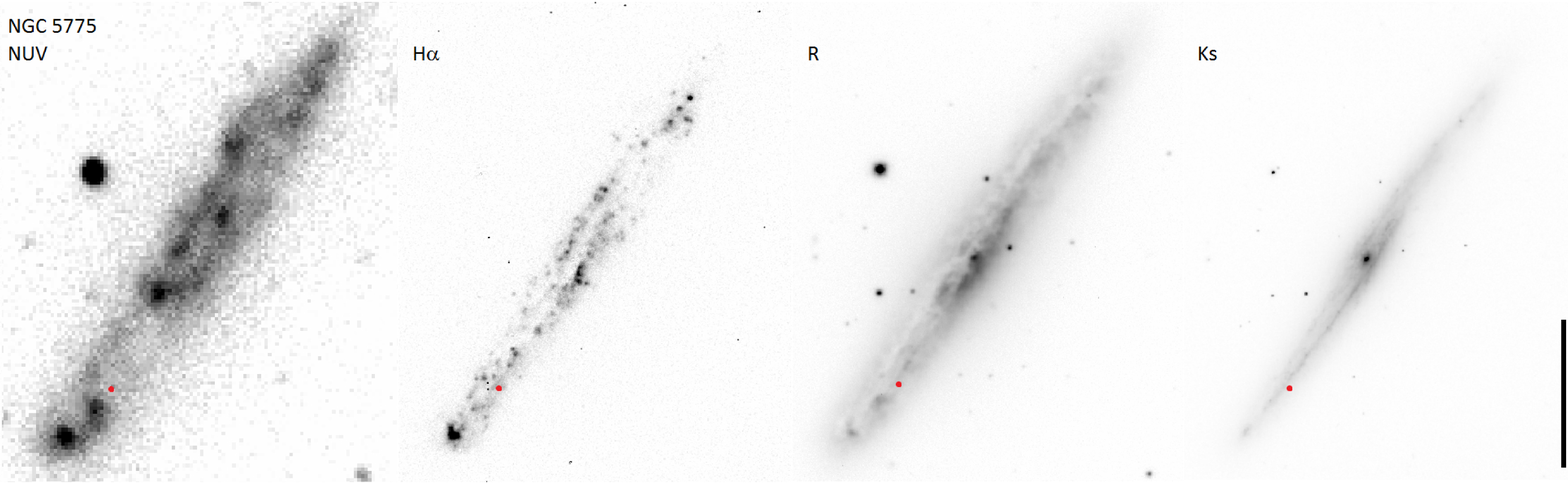}
\includegraphics[width=17cm]{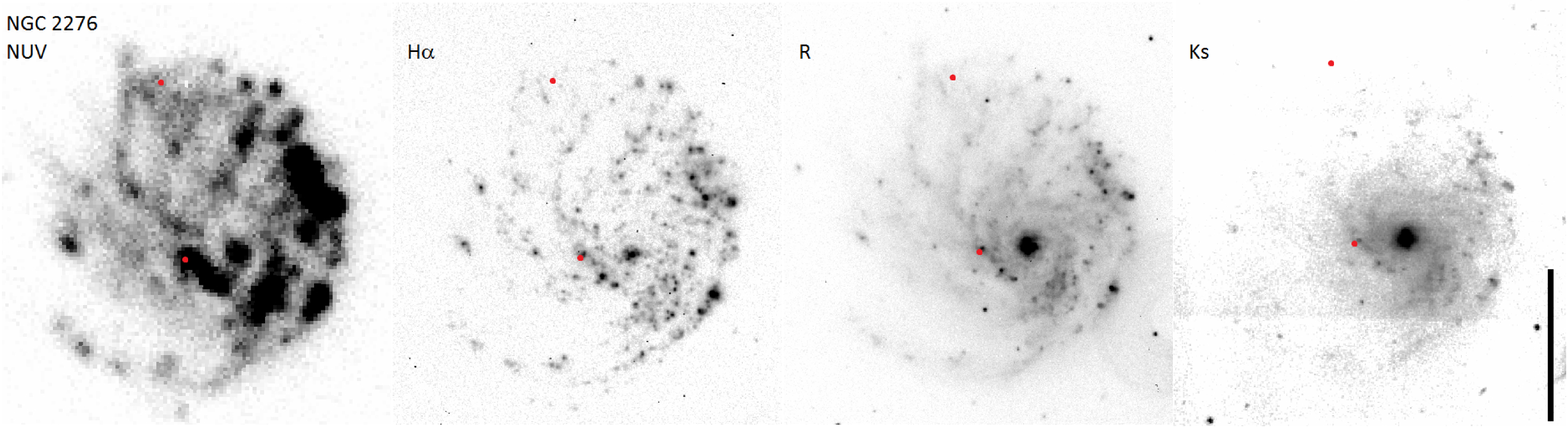}
\end{minipage}
\caption{The datasets of a high- and low-inclination galaxy with typical seeing; NGC 5775 (upper row) and NGC 2276 (lower row). The black line on the right corresponds to one arcminute, and thus roughly 8 kpc in the NGC 5775 images and 10 kpc in the NGC 2276 images. SN positions are marked with red dots.}
\end{figure*}

\section{Data Reduction}

The NOTCam \emph{Ks}-band images were taken with a nine-point (3 $\times$ 3) dither pattern. The sky-subtracted images were de-dithered and median combined. The \emph{Ks}-band images taken with the WHT, dithered between the four quadrants, were reduced using purpose-built scripts that make use of {\sc iraf} tasks. We performed sky subtraction and again median combined the de-dithered images from each epoch into a separate mosaic. Finally, the images of several epochs were combined into a deeper mosaic.

After the basic bias and flat-field corrections of the optical images, further work was done using {\sc iraf} and the \emph{Starlink} {\sc gaia} package \citep{b48}. The H$\alpha$ narrow-band on- and off-filter images and the reduced \emph{Ks}-band images were first aligned with the \emph{R}-band images in {\sc iraf} with the {\sc geomap} and {\sc geotran} tasks, in the {\sc immatch} package, using several field stars identified in each image. Then the continuum was subtracted. The details of this step are described in Subsection 3.1. The world coordinate systems of \emph{R}-band images were fitted by identifying several ($\ge 6$) field stars whose coordinates are listed in the USNO-A2.0 catalog, and performing a coordinate transform using the {\sc iraf} tasks {\sc ccsetwcs} and {\sc ccmap} in the {\sc imcoords} package. Then the coordinate systems were copied to corresponding aligned H$\alpha$ and \emph{Ks}-band images. The RMS errors of the fits were consistently less than 0.5$\arcsec$, with one outlier of 0.6$\arcsec$ and a mean of roughly 0.2$\arcsec$. Assuming that the errors of the reported SN coordinates have a maximum uncertainty of 1.0$\arcsec$, we estimate a conservative combined error in the coordinates to be on the order of 1$\arcsec$, comparable to the median seeing in our data. Images of two galaxies, one with high and one with low inclination, observed with typical seeing, are shown in Fig. 2. The NUV data were pre-reduced and the coordinate systems of the images fitted by the GALEX pipeline.

For each image in each band, background was subtracted; for H$\alpha$ this was done by the reduction script (see below). For the reduced \emph{R}-band, \emph{Ks}-band and NUV images, background was removed by manually measuring the mean background values in the images and subtracting them. 

\subsection{Continuum subtraction}

In order to study the H$\alpha$ emission of the galaxies, the continuum contribution to the flux in the narrow-band H$\alpha$ images was removed. This was done with the {\sc isis} 2.2 package using the Optimal Image Subtraction (OIS) method presented by \citet{b1} and \citet{b2}. In order to subtract the continuum accurately, PSF matching between the on and off images was performed by deriving a space-varying (when needed) convolution kernel. For the kernel derivation, stamps containing a star and some background were identified, including roughly ten stars (if possible) of varying brightness as evenly distributed around the galaxy as possible. For each image pair, a few different sets of options -- mainly stamp size and the degree of spatial variation -- were tested and the best resulting subtracted image (with the most thorough removal of field stars) was chosen. The pixels in the subtracted image were binned by $2 \times 2$ to improve the signal-to-noise ratio.

Our use of this subtraction method contains the intrinsic assumption that the spectral slope (in this case, in the red region) of the selected stars is, on average, similar to that of the target galaxy's stellar population. Despite selecting multiple stars of varying brightness, this may not be entirely true because the image is magnitude-limited and thus a selected stamp star has a higher probability to be bright than if its spectral class was completely random. This should be somewhat compensated for by avoiding the very brightest stars in the fields, which has been done here. 

Although the methods employed here do not require flux calibration, we have tested the reliability of the subtraction by calibrating the H$\alpha$ fluxes of three galaxies -- NGC 3310, NGC 3504 and NGC 7714 -- and checking the values against those reported in the NASA/IPAC Extragalactic Database (NED)\footnote{http://ned.ipac.caltech.edu/}. All three values match within ten per cent. For the flux calibration of these H$\alpha$ images, we used the Isaac Newton Group (ING) list of spectrophotometric standards\footnote{http://www.ing.iac.es/Astronomy/observing/manuals/}. The spectra obtained from the ING website were integrated over the response functions of the NOT filters\footnote{http://www.not.iac.es/instruments/filters/filters.php} using the {\sc iraf} task {\sc sbands}, in the {\sc onedspec} package, to obtain a ``real'' flux value. This was compared to the photometric flux of our standards, measured with \emph{Starlink} {\sc gaia}, to obtain a calibration factor, taking into account also the airmass differences between the target and standard images, and the rebinning.

\section{Analysis methods}

\subsection{Pixel statistics}

The first statistical method we used was correlating the spatial distribution of SNe with the host galaxy emission using pixel statistics. We analysed the sample using the method presented in \citet{b3}, designed to test quantitatively whether SNe of different types follow the H$\alpha$ +[NII] line emission. This method has already been shown to be robust, and its usage allows comparing our results with previous studies. \citet{b3} used a statistical indicator they named the normalised cumulative rank pixel value function (NCRPVF or NCR for short; we adopt the latter abbreviation) to study the sites of a sample of SNe in H$\alpha$. Briefly put, the NCR value of a supernova tells the fraction of the host galaxy emission that originates in regions fainter than the pixel at the SN location (for example, an NCR value of 0.1 means 10 per cent of the galaxy flux comes from fainter pixels and 90 per cent from brighter ones). Thus, this value measures the correlation between SNe and the host galaxy emission. In the case of H$\alpha$ narrow-band images, it measures the correlation between SNe and H II regions (and therefore young massive stars). The NCR distributions of different CCSN types can then be compared.

The H$\alpha$ images were trimmed to include the target galaxy and as little background sky as possible -- the point where the pixel counts reach the background value was chosen as the edge. This usually meant an image of a few hundred $\times$ a few hundred pixels, corresponding to $1-4\arcmin \times 1-4\arcmin$ depending on apparent galaxy size. The pixels of each image were sorted in order of increasing pixel value, giving a sequence from the lowest value (negative) background pixels to the brightest pixels of H$\alpha$ emitting regions. Alongside this sequence the corresponding cumulative distribution was formed, each pixel's respective value obtained by summing the pixel counts in the sequence up to the pixel in question. The cumulative sequence ends in the total H$\alpha$ +[NII] flux of the galaxy. This cumulative rank sequence was then divided by the total line emission flux, and the initial negative NCR values were set to zero, giving a normalised cumulative rank pixel value function running from 0 to 1. The formula for the NCR of a pixel can therefore be expressed as 
\begin{eqnarray}
\mathrm{NCR}_{n} = \sum^{n}_{i=1} P_{i} / \sum^{m}_{j=1} P_{j} \qquad, \mathrm{when} \sum^{n}_{i=1} P_{i} > 0 \\
\mathrm{NCR}_{n} = 0 \qquad, \mathrm{when} \sum^{n}_{i=1} P_{i} \leq 0
\end{eqnarray}
where $n$ is the rank of the pixel whose NCR we want to find within the ascending sequence, $P_{i}$ is the value of a pixel with rank $i$ , and $m$ the total number of pixels. For an illustrative example of the NCR function, see Fig. 6 in \citet{b3}. In our images, the sequence typically contains roughly $10^5$ entries. Finally, based on the coordinates of each SN, the pixel containing the SN was found in the image and assigned an NCR value. 

Before replacing the initial negative values, the NCR reaches a minimum where individual pixel values reach zero. Since the background sky contains roughly equal numbers of positive and negative pixels after subtraction, and, assuming the subtraction has been done accurately, the average value of these pixels is zero, the NCR reaches zero at the point where the sky ends and the emission regions begin, with a value of 1 representing the core of the very brightest H II region or the galactic nucleus. Thus every background pixel has an NCR value of zero in the final sequence, and the choice of how one trims the image has little effect on the result. Since the majority of pixels in H$\alpha$ images are considered background, a population that is randomly scattered throughout the host galaxies tends to low NCR values. One closely associated with the very cores of H II regions would, on the other hand, have high NCR values, while a population accurately following the underlying emission would have a mean value of 0.5 and a uniform NCR distribution (i.e. a distribution where each NCR value has the same probability).

The same steps were repeated for the \emph{R}- and \emph{Ks}-band and NUV images to study the correlations between SNe and galaxy light at those wavelengths as well. 

\subsection{Radial distribution of CCSNe}

Another statistical method we employed was to study the radial distribution of the SNe relative to the flux of the host galaxy.

We determined the fraction of H$\alpha$ +[NII] emission flux -- hereafter called \emph{Fr}(H$\alpha$), as in \citet{b14} -- inside an aperture centered on the \emph{Ks}-band core of the host galaxy. The aperture was chosen in such a way that the location of the SN in question lies on its outer edge. The flux inside this aperture was then divided by the total H$\alpha$ +[NII] flux of the galaxy. Because of the often irregular shape of our galaxies, we used circular apertures instead of the elliptical ones used by e.g. \citet{b3} to compensate for inclination effects. Circular apertures were used for all the galaxies in our sample for the purpose of consistency. These measurements were repeated for the \emph{R}- and \emph{Ks}-band and NUV images. An \emph{Fr} value of 0 would mean that the SN in question has exploded precisely at the core of the galaxy, while a value of 1 means an outlying SN at the very edge of the galaxy (or even outside its visible edge).

There are some issues with the radial distribution analysis. First, it is not trivial to define the ``total flux'' of the galaxy. In this case, we have tried increasing the aperture size until the flux (minus that in the sky annulus) reaches a plateau. Secondly, in some cases it can be difficult to objectively determine where one component of a multiple galaxy system ends and another begins. \citet{b15} attempted to avoid the issue by using the center of the Arp 299 system instead of the cores of individual components, and measuring the flux inside the aperture relative to the whole system flux. We, however, are not sure this kind of method produces truly meaningful results. Another method they used was centering the apertures on the A and B1 cores, as we did, but comparing the aperture flux to the total flux of the system. While this method can be used to study the \emph{relative} \emph{Fr} values in the system (which was the aim of their study), here we cannot obtain results for a sample of galaxies this way. In the case of close galaxy pairs in this study (NGC 3690/IC694, NGC 4038/9 and NGC 4567/8), we used the point of lowest flux between the galaxies as the separating point. The fluxes for both components were measured separately (which produces an underestimated value). The ratio of the measured component fluxes was then used to determine how large a fraction of the total flux of the whole system belongs to which component and the final fluxes of the components calculated from this.

\subsection{Surface density profile of CCSNe}

\citet{b17} derived a scale length for the surface density of SNe in spiral galaxies using 224 SNe within 204 host galaxies. They used a radial distance, corrected for projection effects and normalised to $R_{25}$, the isophotal radius for the blue-band surface brightness of 25 mag arcsec$^{-2}$, to determine a normalised surface density in each distance bin, and fitted an exponential function of the form 
$\Sigma^{SN}(\tilde{r}) = \Sigma^{SN}_0$exp$(-\tilde{r}/\tilde{h}_{SN})$. Here $\tilde{r}$ is the distance normalised to $R_{25}$ and $\tilde{h}_{SN}$ is the scale length of the surface density, also normalised to $R_{25}$. Herrero-Illana, P\'{e}rez-Torres \& Alberdi (2012) applied the same method to VLBI-observed SN remnants and radio SNe in a starburst galaxy, a LIRG and an ULIRG. We similarly determine a normalised surface density distribution for our sample of CCSNe.

We calculated the de-projected distance of each SN to the center of its host galaxy according to equations (1) and (2) in \citet{b17}:
\begin{equation}
U = \Delta\alpha \mathrm{sin PA} + \Delta\delta \mathrm{cos PA}, V = \Delta\alpha \mathrm{cos PA} - \Delta\delta \mathrm{sin PA}
\end{equation}
\begin{equation}
R_{SN}^2 = U^2 + \left(\frac{V}{\cos i}\right)^2
\end{equation}
where U and V are the coordinates of the SN in the host galaxy's coordinate system, PA is the major axis position angle, $\Delta\alpha$ and $\Delta\delta$ are the coordinate differences between the SN and the galaxy core, $i$ the galaxy inclination and $R_{SN}$ the true SN-core distance. However, during our analysis, it became clear that as the inclination of the host galaxy approaches 90$^{\circ}$, the calculated distance becomes too large to be trustworthy. Thus, galaxies with inclination $> 70^{\circ}$ were excluded from this study, leaving us with 77 SNe in 49 galaxies.

The de-projected distances were normalised to the $R_{25}$ of the host galaxy. The coordinates of the cores were determined from the \emph{Ks}-band images for the galaxies for which we had those images available (all but one), and taken from the NED for NGC 6951. The $R_{25}$ distances, position angles of the major axes of the galaxies and inclinations of the galactic disks were obtained from the HyperLeda database\footnote{http://leda.univ-lyon1.fr/} \citep{b33} if available. If the HyperLeda database did not have the information for a particular galaxy, we used NED instead. We then divided the normalised distance distribution into quartile bins. The surface density in each bin was determined by dividing the number of SNe in the bin by the area of a concentric ring (normalised to $R_{25}^2$), with the limits of the bin as the inner and outer edges of the ring, around the nucleus. The unit of surface density is thus $R_{25}^{-2}$.

We fitted two different disk profiles, an exponential disk of the form $\Sigma^{SN}(\tilde{r}) = \Sigma^{SN}_0$exp$(-\tilde{r}/\tilde{h}_{SN})$  and a disk with power-law surface density profile $\Sigma^{SN}(\tilde{r}) = \Sigma^{SN}_0\tilde{r}^{-\gamma}$, to the data. The first case assumes that the SN distribution follows the radial surface brightness distribution in spiral disks \citep{b18}; while the second has been used to probe profiles predicted by numerical simulations \citep{b19}. Since our sample sizes in each bin are relatively small, we performed a Monte Carlo (MC) simulation to obtain more reliable values for the fit parameters, similarly to \citet{b23}. We generated 10000 mock samples whose surface density values in each bin are uniformly distributed within the Poissonian uncertainty of each data point. We then fitted the exponential and power-law functions to each of these samples to obtain 10000 sets of parameters. We then used the median values of the parameters $\gamma$ and $\tilde{h}_{SN}$ from the MC distributions. The uncertainty in these values was set at the 90 per cent confidence level.

\section{Results}

\subsection{Pixel statistics}

The NCR values based on the H$\alpha$ +[NII] line, \emph{R}- and \emph{Ks}-band and NUV light are listed in Table 2, along with the results for the radial distribution (see below), for the 86 CCSNe. We have grouped our CCSNe into two subsamples, those of type Ibc/IIb -- thought to have stripped envelope progenitors -- and the remaining type II SNe. In Table 3 we present the mean NCR values of all subsamples, along with the standard errors of the mean and the number of SNe in the different groups. In this table and others, the Ibc subsample also includes the SNe classified as Ib/c, which are not included in the Ib or Ic groupings. The distributions of SNe are presented on the left-hand side of Fig. 3 as cumulative distribution functions (CDFs). The mean NCR(H$\alpha$) value ($\pm$ SEM) of type Ibc/IIb SNe is statistically significantly larger (0.475 $\pm$ 0.052) than for type II (0.268 $\pm$ 0.038). Further separating the type Ibc sample into Ib and Ic subsamples, we find a mean NCR of 0.607 $\pm$ 0.068 for Ic as opposed to 0.290 $\pm$ 0.074 for Ib, implying type Ic having a stronger association with H II regions than Ib -- and that, within the error limits, the correlations for types II and Ib are roughly equal. In the \emph{R}-band, the difference between the average values for the two types is somewhat less (0.619 $\pm$ 0.041 for type Ibc/IIb vs. 0.516 $\pm$ 0.041 for type II). The NCR(\emph{Ks}) values are closer to the ones from H$\alpha$ images, with a mean of 0.499 $\pm$ 0.041 for type Ibc/IIb and 0.394 $\pm$ 0.043 for type II. Finally, in the NUV, the difference between the two types is just within the error: 0.614 $\pm$ 0.052 for type Ibc/IIb and 0.525 $\pm$ 0.047 for type II. For subtypes of type I, we get 0.482 $\pm$ 0.112 for type Ib and 0.666 $\pm$ 0.072 for Ic; the difference is just barely larger than the error. This indicates type Ib events more closely follow the host galaxy's NUV light distribution, while type Ic SNe preferably occur in the brighter regions.

\citet{b21} obtained a mean NCR(H$\alpha$) value of 0.254 $\pm$ 0.023 for type II SNe, which is consistent with our value for type II. For type Ibc, they obtained a mean value of 0.390 $\pm$ 0.031, which is significantly less than our corresponding value, 0.500 $\pm$ 0.055.

We used the Kolmogorov-Smirnov (K-S) test to compare our distributions to a hypothetical flat (i.e. uniform) distribution. All results for the K-S tests are listed in Table 5. Since the NCR tells how large a fraction of the galaxy's flux comes from pixels fainter than the SN's location, a similar fraction of CCSNe should also explode in fainter pixels. Therefore, if the SNe accurately follow the H$\alpha$ +[NII] line emission, the distribution of NCR values should be flat, with a mean of 0.5. Thus, this comparison tells us how strongly each sub-sample, or the whole population of CCSNe, is correlated with the star-forming regions. We found that the distribution of the NCR values of type Ibc/IIb SNe is formally consistent with a flat distribution -- and the mean NCR value is indeed consistent with 0.5 within the error. The NCRs of type II SNe, on the other hand, do differ from a flat distribution, with the probability of the distributions being the same being under 0.1 per cent. Thus, type Ibc/IIb SNe do seem to follow the H$\alpha$ +[NII] line emission rather well, while type II SNe do not.

\begin{figure*}
\centering
\begin{minipage}{160mm}
\begin{tabular}[t]{cc}
\includegraphics[width=8cm]{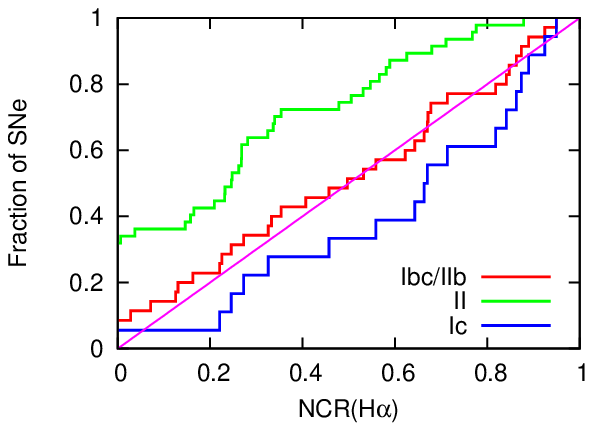} & \includegraphics[width=8cm]{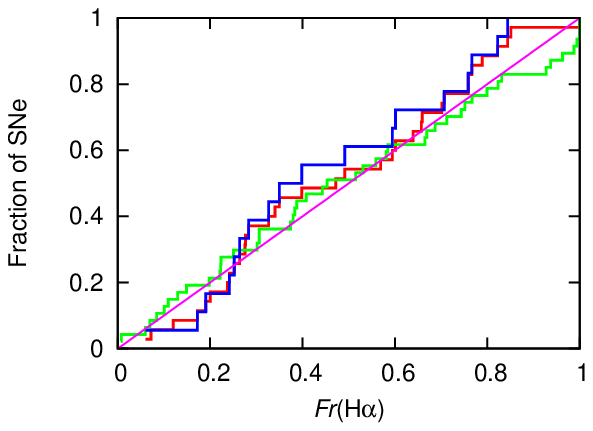} \\
\includegraphics[width=8cm]{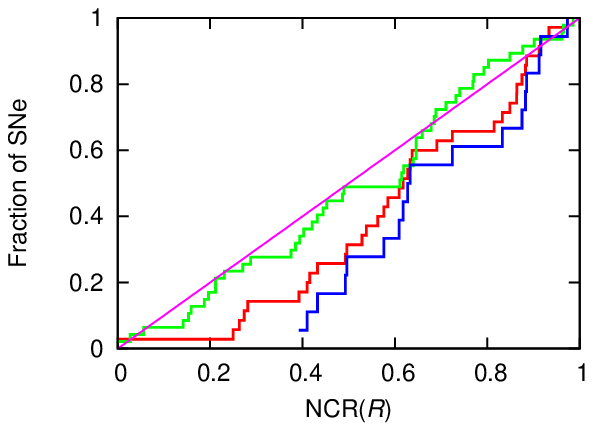} & \includegraphics[width=8cm]{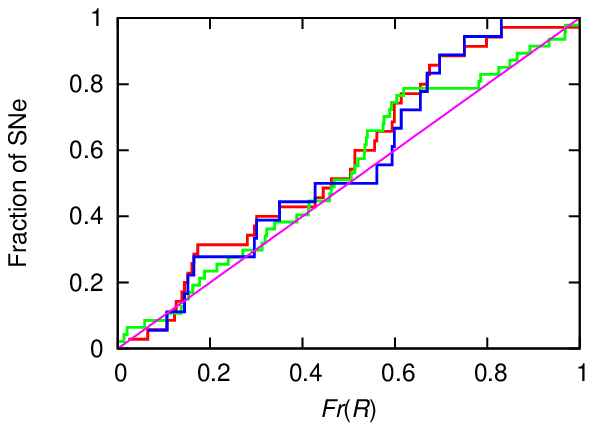} \\
\includegraphics[width=8cm]{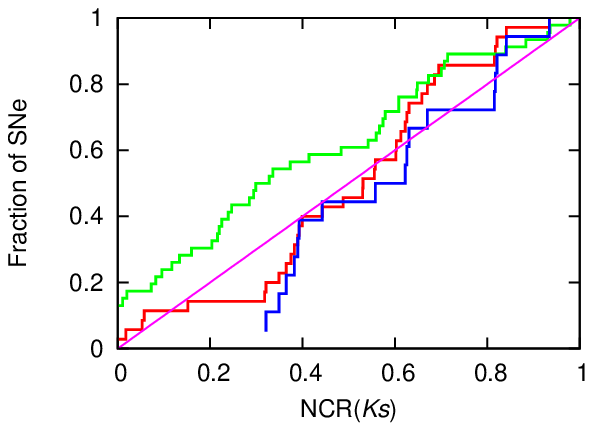} & \includegraphics[width=8cm]{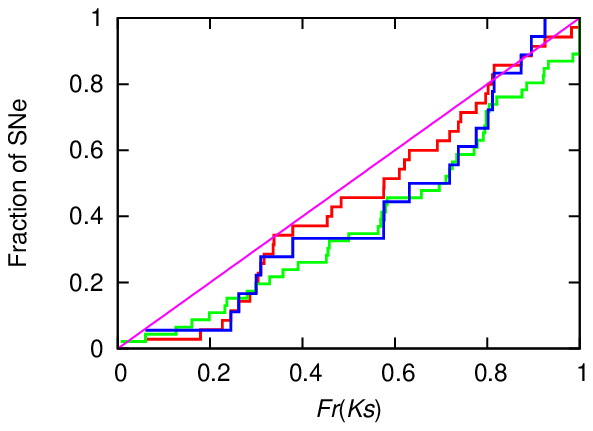} \\
\includegraphics[width=8cm]{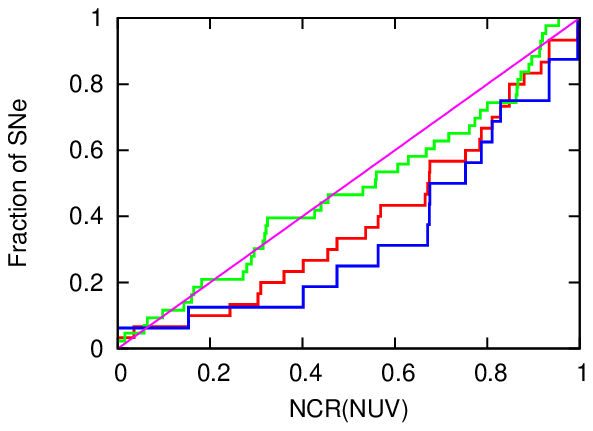} & \includegraphics[width=8cm]{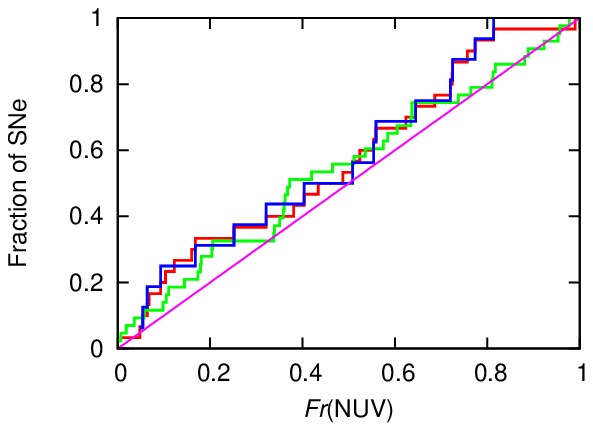} \\
\end{tabular}
\caption{The cumulative distributions of NCR (left) and \emph{Fr} (right) values for the SNe of different types in different bands. The red lines represent the type Ibc/IIb subsample, the blue lines just the type Ic and the green lines the type II subsample, while the purple diagonal line is the CDF for a flat distribution.}
\end{minipage}
\end{figure*}

\begin{table*}
\begin{minipage}{170mm}
\centering
\caption{The main results for the CCSN sample from $H\alpha$, \emph{R}- and \emph{Ks}-band and NUV images.}
\begin{tabular}[t]{lcccccccccc}
    \hline
         SN & Type & Host galaxy & NCR(H$\alpha$) & NCR(\emph{R}) & NCR(\emph{Ks}) & NCR(NUV) & \emph{Fr}(H$\alpha$) & \emph{Fr}(\emph{R}) & \emph{Fr}(\emph{Ks}) & \emph{Fr}(NUV)\\
    \hline
    2012es & IIb & NGC 5597 & 0.125 & 0.538 & 0.375 & 0.666 & 0.640 & 0.463 & 0.621 & 0.434 \\
    2011dh & IIb & M 51 & 0.028 & 0.274 & 0.053 & 0.569 & 0.659 & 0.798 & 0.464 & 0.757 \\
    2010gk & Ic & NGC 5433 & 0.818 & 0.833 & 0.622 & 0.934 & 0.327 & 0.614 & 0.576 & 0.168 \\
    2010bk & II P & NGC 4433 & 0.589 & 0.711 & 0.840 & 0.761 & 0.101 & 0.058 & 0.160 & 0.036 \\
    2010as & Ib/c & NGC 6000 & 0.678 & 0.815 & 0.686 & - & 0.238 & 0.160 & 0.336 & -\\
    2010P & Ib/IIb & NGC 3690 & 0.354 & 0.563 & 0.488 & 0.537 & 0.789 & 0.504 & 0.609 & 0.487 \\
    2010O & Ib & IC 694 & 0.497 & 0.864 & 0.603 & 0.455 & 0.277 & 0.123 & 0.286 & 0.103 \\
    2009jf & Ib & NGC 7479 & 0.532 & 0.282 & 0.152 & 0.880 & 0.569 & 0.675 & 0.796 & 0.524\\
    2009hd & II & M 66 & 0.776 & 0.619 & 0.579 & 0.685 & 0.306 & 0.413 & 0.569 & 0.362\\
    2009ga & II P & NGC 7678 & 0.210 & 0.646 & 0.484 & 0.426 & 0.712 & 0.589 & 0.719 & 0.605\\
    2009em & Ic & NGC 157 & 0.457 & 0.576 & 0.390 & 0.671 & 0.283 & 0.350 & 0.576 & 0.321\\
    2009H & II & NGC 1084 & 0.354 & 0.385 & 0.294 & 0.324 & 0.585 & 0.593 & 0.795 & 0.536\\
    2008iz & II & M 82 & 0.337 & 0.793 & 0.936 & 0.915 & 0.084 & 0.014 & 0.126 & 0.007\\
    2008in & II P & M 61 & 0 & 0.188 & 0.011 & 0.160 & 0.937 & 0.850 & 0.932 & 0.810 \\
    2008ij & II & NGC 6643 & 0.479 & 0.421 & 0.225 & - & 0.383 & 0.322 & 0.696 & -\\
    2008ho & II P & NGC 922 & 0.532 & 0.445 & 0.117 & 0.668 & 0.928 & 0.892 & 0.922 & 0.881\\
    2008gz & II & NGC 3672 & 0.232 & 0.877 & 0.649 & 0.913 & 0.109 & 0.162 & 0.198 & 0.057 \\
    2008eb & Ic & NGC 6574 & 0.273 & 0.410 & 0.318 & 0.154 & 0.767 & 0.594 & 0.631 & 0.645\\
    2008co & - & IC 2522 & 0.693 & 0.915 & 0.879 & 0.705 & 0.028 & 0.057 & 0.094 & 0.043 \\
    2008br & II & IC 2522 & 0 & 0.640 & 0.285 & 0.717 & 0.060 & 0.153 & 0.237 & 0.206 \\
    2008bp & II P & NGC 3095 & 0 & 0.288 & 0 & 0.279 & 0.221 & 0.387 & 0.583 & 0.584 \\
    2008bo & Ib & NGC 6643 & 0 & 0.637 & 0.530 & - & 0.657 & 0.513 & 0.982 & -\\
    2007fo & Ib & NGC 7714 & 0.226 & 0.529 & 0.556 & 0.304 & 0.702 & 0.556 & 0.454 & 0.624 \\
    2007ch & II & NGC 6000 & 0.146 & 0.394 & 0.220 & - & 0.752 & 0.577 & 0.798 & -\\
    2007aa & II P & NGC 4030 & 0.157 & 0.056 & 0 & 0.318 & 0.963 & 0.968 & 0.985 & 0.953 \\
    2006ov & II P & M 61 & 0.165 & 0.771 & 0.647 & 0.890 & 0.387 & 0.459 & 0.571 & 0.358 \\
    2006gi & Ib & NGC 3147 & 0 & 0 & 0.018 & 0.035 & 1 & 1 & 1 & 0.990 \\
    2006A & - & NGC 7753 & 0.139 & 0.707 & 0.754 & 0.549 & 0.060 & 0.238 & 0.186 & 0.072\\
    2005kh & II & NGC 3094 & 0 & 0 & 0.02 & 0.016 & 1 & 1 & 1 & 0.889\\
    2005lr & Ic & ESO 492-G02 & 0.246 & 0.493 & 0.321 & - & 0.759 & 0.427 & 0.802 & -\\
    2005eb & II & UGC 556 & 0.680 & 0.611 & 0.239 & 0.800 & 0.744 & 0.535 & 0.772 & 0.360 \\
    2005dl & II & NGC 2276 & 0.768 & 0.849 & 0.714 & 0.863 & 0.070 & 0.105 & 0.391 & 0.145\\
    2005cs & II P & M 51 & 0.505 & 0.741 & 0.574 & 0.896 & 0.224 & 0.312 & 0.232 & 0.204 \\
    2005V & Ib/c & NGC 2146 & 0.848 & 0.934 & 0.695 & 0.917 & 0.072 & 0.023 & 0.179 & 0.004\\
    2005U & IIb & IC 694 & 0.622 & 0.849 & 0.613 & 0.848 & 0.341 & 0.173 & 0.338 & 0.160 \\
    2005H & II & NGC 838 & 0.566 & 0.685 & 0.560 & 0.785 & 0.306 & 0.188 & 0.280 & 0.105\\
    2004gt & Ic & NGC 4038 & 0.863 & 0.974 & 0.925 & 0.996 & 0.844 & 0.655 & 0.812 & 0.725\\
    2004gn & Ic & NGC 4527 & 0.663 & 0.633 & 0.443 & 0 & 0.350 & 0.599 & 0.925 & 0.404 \\
    2004ej & II & NGC 3095 & 0 & 0.142 & 0.073 & 0.289 & 0.130 & 0.271 & 0.452 & 0.368 \\
    2004cc & Ic & NGC 4568 & 0.714 & 0.912 & 0.841 & 0.564 & 0.190 & 0.165 & 0.262 & 0.064 \\
    2004bf & Ic & UGC 8739 & 0.874 & 0.885 & 0.393 & 0.787 & 0.492 & 0.750 & 0.873 & 0.559 \\
    2004am& II P & M 82 & 0.547 & 0.965 & 0.884 & 0.606 & 0.582 & 0.240 & 0.564 & 0.098\\
    2004C & Ic & NGC 3683 & 0.950 & 0.618 & 0.626 & 0.676 & 0.707 & 0.670 & 0.896 & 0.508 \\
    2003iq & II & NGC 772 & 0 & 0.402 & 0.160 & 0.558 & 0.374 & 0.534 & 0.885 & 0.338\\
    2003hl & II & NGC 772 & 0.262 & 0.689 & 0.414 & 0.773 & 0.198 & 0.340 & 0.657 & 0.174\\
    2003hg & II & NGC 7771 & 0.246 & 0.659 & 0.673 & 0.556 & 0.453 & 0.215 & 0.455 & 0.351\\
    2003ao & II P & NGC 2993 & 0.037 & 0.487 & 0.374 & 0.531 & 0.531 & 0.319 & 0.358 & 0.637 \\
    2003B & II P & NGC 1097 & 0 & 0.027 & 0 & 0.064 & 1 & 0.969 & 1 & 0.812\\
    2002ji & Ib/c & NGC 3655 & 0.131 & 0.250 & 0.057 & 0.360 & 0.472 & 0.513 & 0.692 & 0.724 \\
    2002gw & II & NGC 922 & 0 & 0.211 & 0.204 & 0.182 & 0.799 & 0.825 & 0.820 & 0.764\\
    2001is & Ib & NGC 1961 & 0.333 & 0.263 & 0 & 0.784 & 0.200 & 0.281 & 0.743 & 0.381\\
    2001ej & Ib & UGC 3829 & 0.163 & 0.585 & 0.531 & - & 0.851 & 0.445 & 0.317 & -\\
    2001ci & Ic & NGC 3079 & 0.221 & 0.725 & 0.671 & 0.475 & 0.242 & 0.145 & 0.310 & 0.093 \\
    2001ac & IIn & NGC 3504 & 0 & 0.153 & 0.082 & 0 & 0.988 & 0.865 & 0.875 & 0.923\\
    2000cr & Ic & NGC 5395 & 0.925 & 0.627 & 0.558 & 0.829 & 0.822 & 0.831 & 0.719 & 0.774 \\
    2000C & Ic & NGC 2415 & 0.670 & 0.392 & 0.631 & 0.675 & 0.399 & 0.301 & 0.737 & 0.720\\
    1999gn & II P & M 61 & 0.581 & 0.769 & 0.542 & 0.873 & 0.408 & 0.469 & 0.580 & 0.372 \\
    1999gl & II & NGC 317B & 0.340 & 0.732 & 0.609 & 0.920 & 0.687 & 0.145 & 0.329 & 0.110\\ 
    1999eu & II P & NGC 1097 & 0 & 0.159 & 0 & 0.143 & 0.994 & 0.934 & 1 & 0.738\\
    1999el & IIn & NGC 6951 & 0 & 0.646 & 0.551 & 0.440 & 0.251 & 0.137 & 0.646 & 0.179\\
    1999dn & Ib & NGC 7714 & 0.072 & 0.416 & 0.399 & 0.243 & 0.759 & 0.599 & 0.484 & 0.686 \\
    1999bx & II & NGC 6745 & 0.268 & 0.646 & 0.299 & - & 0.380 & 0.462 & 0.299 & -\\
    1999cz & Ic & NGC 5078 & 0 & 0.609 & 0.365 & 0.811 & 0.595 & 0.561 & 0.776 & 0.814 \\
\hline
\end{tabular}
\end{minipage}
\end{table*}
\addtocounter{table}{-1}
\begin{table*}
\begin{minipage}{170mm}
\centering
\caption{The main results for the CCSN sample from $H\alpha$, \emph{R}- and \emph{Ks}-band and NUV images (continued).}
\begin{tabular}[t]{lcccccccccc}
    \hline
         SN & Type & Host galaxy & NCR(H$\alpha$) & NCR(\emph{R}) & NCR(\emph{Ks}) & NCR(NUV) & \emph{Fr}(H$\alpha$) & \emph{Fr}(\emph{R}) & \emph{Fr}(\emph{Ks}) & \emph{Fr}(NUV)\\
    \hline
    1999D & II & NGC 3690 & 0 & 0.271 & 0.134 & 0.165 & 1 & 0.785 & 0.791 & 0.978 \\
    1998dl & II P & NGC 1084 & 0.268 & 0.375 & 0.216 & 0.455 & 0.559 & 0.575 & 0.778 & 0.512\\
    1998cf & - & NGC 3504 & 0.915 & 0.972 & 0.979 & 0.925 & 0.031 & 0.016 & 0.028 & 0.011\\
    1998Y & II & NGC 2415 & 0.248 & 0.679 & 0.567 & 0.866 & 0.223 & 0.178 & 0.500 & 0.465\\
    1998T & Ib & IC 694 & 0.672 & 0.864 & 0.658 & 0.848 & 0.275 & 0.139 & 0.304 & 0.123 \\
    1997dc & Ib & NGC 7678 & 0.407 & 0.691 & 0.602 & 0.310 & 0.120 & 0.127 & 0.227 & 0.068\\
    1997bs\footnote{SN 1997bs has been considered a SN ``impostor'' by \citet{b34}, but there is also evidence that this is not the case \citep{b35}. Thus we decided to keep it in the sample.} & IIn & M 66 & 0.281 & 0.452 & 0.336 & 0.315 & 0.669 & 0.538 & 0.725 & 0.637\\
    1996W & II & NGC 4027 & 0.879 & 0.803 & 0.707 & - & 0.767 & 0.540 & 0.797 & \\
    1996an & II & NGC 1084 & 0.268 & 0.902 & 0.701 & 0.955 & 0.443 & 0.506 & 0.710 & 0.419\\
    1996ae & IIn & NGC 5775 & 0.626 & 0.490 & 0.328 & 0.271 & 0.665 & 0.782 & 0.923 & 0.575 \\
    1996D & Ic & NGC 1614 & 0.326 & 0.433 & 0.383 & - & 0.264 & 0.065 & 0.300 &\\
    1994ak & IIn & NGC 2782 & 0 & 0.196 & 0 & 0.321 & 0.831 & 0.52 & 1 & 0.818\\
    1994Y & IIn & NGC 5371 & 0 & 0.614 & 0.609 & 0.057 & 0.149 & 0.414 & 0.458 & 0.181 \\
    1994W\footnote{The SN nature of SN 1994W has been questioned \citep{b45}; however, contrary evidence for alike transients has also been found, see e.g. the discussion of \citet{b46}. Thus we decided to keep it in the sample as well.} & IIn & NGC 4041 & 0.232 & 0.432 & 0.247 & 0.629 & 0.302 & 0.513 & 0.804 & 0.340\\
    1994I & Ic & M 51 & 0.643 & 0.875 & 0.821 & 0.934 & 0.061 & 0.107 & 0.060 & 0.048\\
    1993X & II & NGC 2276 & 0.007 & 0.231 & 0 & 0.295 & 0.515 & 0.619 & 1 & 0.957\\
    1993G & II L & IC 694 & 0 & 0.212 & 0.095 & 0.097 & 0.823 & 0.604 & 0.734 & 0.634 \\
    1992bu & - & NGC 3690 & 0.227 & 0.675 & 0.471 & 0.538 & 0.309 & 0.261 & 0.434 & 0.210\\
    1992bd & II & NGC 1097 & 0.325 & 0.986 & 0.979 & 0.865 & 0.008 & 0.003 & 0.007 & 0.001\\
    1991N & Ic & NGC 3310 & 0.890 & 0.916 & 0.818 & 0.996 & 0.252 & 0.296 & 0.379 & 0.252\\
    1991J & II & NGC 5020 & 0.711 & 0.961 & 0.931 & 0.927 & 0.007 & 0.020 & 0.060 & 0.019 \\
    1990U & Ic & NGC 7479 & 0.559 & 0.496 & 0.349 & 0.753 & 0.601 & 0.697 & 0.815 & 0.554\\
    1990B & Ic & NGC 4568 & 0.842 & 0.883 & 0.815 & 0.402 & 0.173 & 0.152 & 0.245 & 0.054 \\
\hline
\end{tabular}
\end{minipage}
\end{table*}

\begin{table*}
\centering
\begin{minipage}{135mm}
\caption{The mean H$\alpha$, \emph{R}- and \emph{Ks}-band and NUV NCR values of the CCSNe, with comparison values from A12.}
\begin{tabular}[t]{lcccccc}
    \hline
         SN type & N & NCR(H$\alpha$) & NCR(H$\alpha$, A12) & NCR(\emph{R}) & NCR(\emph{Ks}) & NCR(NUV)\\
    \hline
    \hline
	Ibc/IIb & 35 & 0.475 $\pm$ 0.052 & - & 0.619 $\pm$ 0.041 & 0.499 $\pm$ 0.041 & 0.614 $\pm$ 0.052 \\
	II & 47 & 0.268 $\pm$ 0.038 & 0.254 $\pm$ 0.023 & 0.516 $\pm$ 0.041 & 0.394 $\pm$ 0.043 & 0.525 $\pm$ 0.047\\
    \hline
	Ibc & 31 & 0.500 $\pm$ 0.055 & 0.390 $\pm$ 0.031 & 0.627 $\pm$ 0.044 & 0.513 $\pm$ 0.044 & 0.607 $\pm$ 0.059 \\
	Ib & 10 & 0.290 $\pm$ 0.074 & 0.318 $\pm$ 0.045 & 0.515 $\pm$ 0.088 & 0.418 $\pm$ 0.084 & 0.482 $\pm$ 0.112 \\
	Ic & 18 & 0.607 $\pm$ 0.068 & 0.469 $\pm$ 0.040 & 0.683 $\pm$ 0.046 & 0.572 $\pm$ 0.049 & 0.666 $\pm$ 0.072\\
    \hline
	All & 86 & 0.363 $\pm$ 0.033 & - & 0.573 $\pm$ 0.029 & 0.454 $\pm$ 0.031 & 0.567 $\pm$ 0.034 \\
\hline
\end{tabular}
\end{minipage}
\end{table*}

\begin{table*}
\centering
\begin{minipage}{132mm}
\caption{The mean H$\alpha$, \emph{R}- and \emph{Ks}-band and NUV \emph{Fr} values of the CCSNe, with comparison values (marked H12) for the SNe from \citet{b31} and Stacey Habergham, private communication.}
\begin{tabular}[t]{lcccccc}
    \hline
         SN type & N & \emph{Fr}(H$\alpha$) & \emph{Fr}(H$\alpha$, H12) & \emph{Fr}(\emph{R}) & \emph{Fr}(\emph{Ks}) & \emph{Fr}(NUV)\\
    \hline
    \hline
	Ibc/IIb & 35 & 0.488 $\pm$ 0.044 & 0.465 $\pm$ 0.024 & 0.438 $\pm$ 0.043 & 0.561 $\pm$ 0.046 & 0.426 $\pm$ 0.052 \\
	II & 47 & 0.502 $\pm$ 0.046 & 0.552 $\pm$ 0.027 & 0.469 $\pm$ 0.041 & 0.607 $\pm$ 0.042 & 0.453 $\pm$ 0.047\\
    \hline
	Ibc & 31 & 0.472 $\pm$ 0.048 & 0.456 $\pm$ 0.026 & 0.433 $\pm$ 0.046 & 0.570 $\pm$ 0.051 & 0.420 $\pm$ 0.058\\
	Ib & 10 & 0.575 $\pm$ 0.088 & 0.524 $\pm$ 0.047 & 0.474 $\pm$ 0.087 & 0.578 $\pm$ 0.102 & 0.437 $\pm$ 0.117\\
	Ic & 18 & 0.451 $\pm$ 0.059 & 0.431 $\pm$ 0.037 & 0.443 $\pm$ 0.059 & 0.594 $\pm$ 0.063 & 0.419 $\pm$ 0.070\\
    \hline
	All & 86 & 0.478 $\pm$ 0.032 & 0.503 $\pm$ 0.018 & 0.441 $\pm$ 0.029 & 0.569 $\pm$ 0.031 & 0.423 $\pm$ 0.034\\
\hline
\end{tabular}
\end{minipage}
\end{table*}

\begin{table*}
\centering
\begin{minipage}{147mm}
\caption{Results of the K-S tests between different samples and between the samples and a flat distribution, given as probabilities that the compared distributions are the same. When a difference is significant at the 90 per cent confidence level, the corresponding value is written in bold.}
\begin{tabular}[t]{lccccccc}
  \hline
	Distr. & Ibc/IIb vs. flat & II vs. flat & Ibc/IIb vs. II& Ib vs. flat & {Ib vs. II} & Ic vs. flat & Ic vs. II \\
  \hline
	NCR(H$\alpha$) & 95 \% & {\bf $<$ 0.1 \%} & {\bf 2.7 \%} & 10 \% & 97 \% & 43 \% & {\bf $<$ 1 \%} \\
	NCR(\emph{R}) & {\bf 5.6 \%} & 71 \% & 20 \% & $>$ 99 \% & 99 \% & {\bf 2.5 \%} & 11 \% \\
	NCR(\emph{Ks}) & 29 \% & {\bf 4.7 \%} & {\bf 1.2 \%} & 15 \% & 60 \% & 14 \% & \textbf{$<$ 1 \%} \\
	NCR(NUV) & 16 \% & 81 \% & 47 \% & 90 \% & 93 \% & {\bf 9.9 \%} & 23 \% \\
	\emph{Fr}(H$\alpha$) & 66 \% & 94 \% & 78 \% & $>$ 99 \% & 60 \% & 73 \% & 81 \% \\
	\emph{Fr}(\emph{R}) & 15 \% & 13 \% & 88 \% & 63 \% & 98 \% & 46 \% & 69 \% \\
	\emph{Fr}(\emph{Ks}) & 46 \% & {\bf 8.2 \%} & 70 \% & $>$ 99 \% & 57 \% & 51 \% & 91 \% \\
	\emph{Fr}(NUV) & 35 \% & 35 \% & 84 \% & 62 \% & 95 \% & 59 \% & 89 \% \\
  \hline
\end{tabular}
\end{minipage}
\end{table*}

\begin{figure*}
\centering
\begin{minipage}{87mm}
\includegraphics[width=8.7cm]{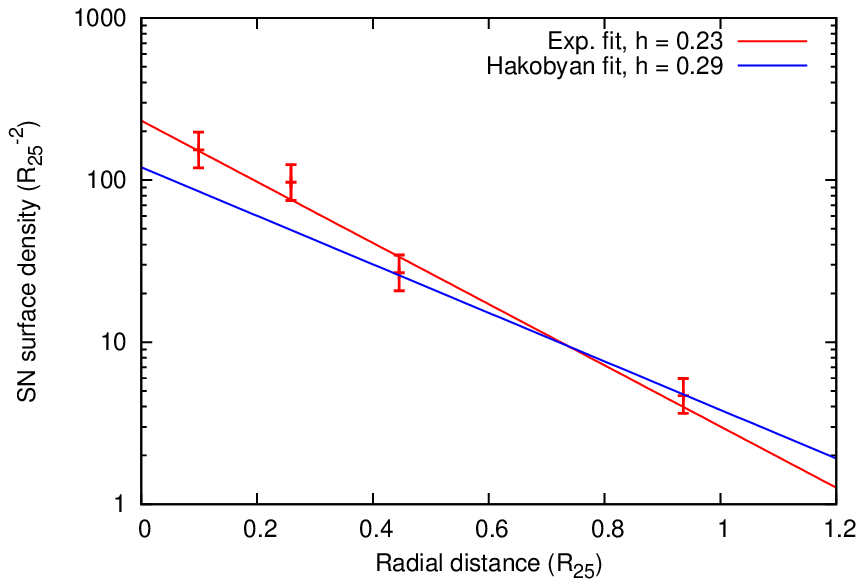}
\end{minipage}
\begin{minipage}{87mm}
\includegraphics[width=8.7cm]{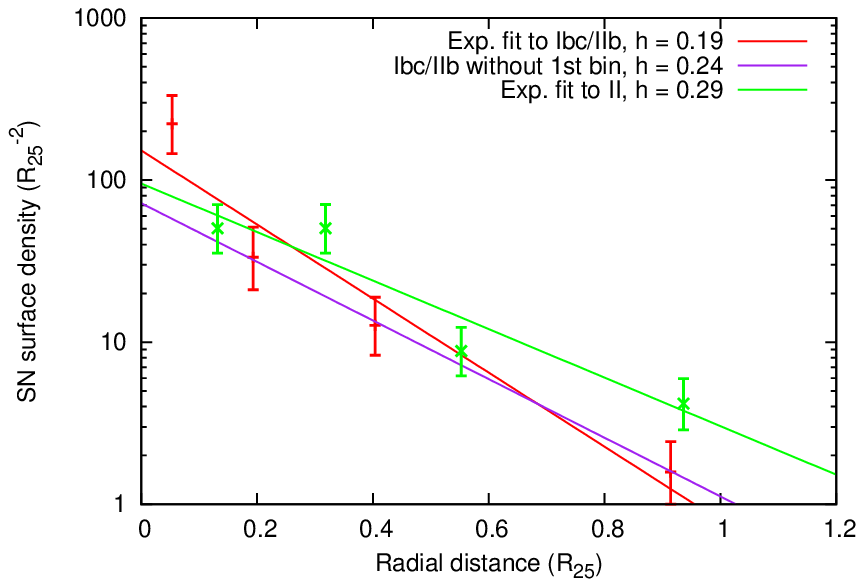}
\end{minipage}
\caption{The SN surface density values as a function of radial distance (Points with error bars) and exponential fits derived from our Monte Carlo analysis (straight lines). The full sample is on the left and a comparison between subsamples is presented on the right (red for types Ibc/IIb and green for type II). The scale length derived by \citet{b17} only fits the outer regions for our sample, while for fitting the full sample a shorter scale length is required. The scale length for types Ibc/IIb is also shorter than for type II, although the difference is somewhat mitigated if one ignores the most central radial distance bin.}
\end{figure*}

We repeated this test for the NCR values from the other bands, and found that, in the \emph{R}-band, the Ibc/IIb sample significantly differs from a flat distribution while type II does not. However, the type Ib sample seems to trace the galaxy light as well. In the \emph{Ks}-band, the results are similar to H$\alpha$, with a statistically significant difference from a flat distribution for type II but not for type Ibc/IIb, but the test cannot detect differences between types Ib and Ic. In the NUV, only the Ic sample significantly differs from a flat distribution. With a mean value very close to 0.5, however, type II follows the galaxy emission more accurately than types Ibc/IIb.

Furthermore, we performed two-sample K-S tests to compare the distributions of the different types. The difference between the distributions of types Ibc/IIb and II is significant in H$\alpha$ and in the \emph{Ks}-band -- and subtypes Ib and Ic also show a significant difference in H$\alpha$. This agrees quite well with the differences in mean NCR values. Further two-sample tests were performed between the H$\alpha$ and NUV values, to see if the distributions match regarding recent and on-going star formation. The only significant difference here is for type II.

\subsection{Radial distribution of CCSNe}

The \emph{Fr} values for all wavelengths are listed in Table 2, and their cumulative distributions are presented on the right-hand side of Fig. 3. The mean values and standard errors of the mean are listed in Table 4. The radial distributions of SNe of different types do not have a difference as big as the NCR values: the mean \emph{Fr}(H$\alpha$) for type Ibc/IIb is 0.488 $\pm$ 0.044, while for type II SNe it is 0.502 $\pm$ 0.046. For subtypes Ib and Ic the values are 0.575 $\pm$ 0.088 and 0.451 $\pm$ 0.059, respectively. In the \emph{R}-band the values are consistent with those in H$\alpha$ within the errors: 0.438 $\pm$ 0.043 for Ibc/IIb, 0.474 $\pm$ 0.087 for Ib, 0.443 $\pm$ 0.059 for Ic and 0.469 $\pm$ 0.041 for type II. In the \emph{Ks}-band we see a difference to the H$\alpha$ and \emph{R}-band, with a mean value of 0.561 $\pm$ 0.044 for types Ibc/IIb and 0.607 $\pm$ 0.042 for type II; while the NUV values are close to the ones from \emph{R}-band, 0.426 $\pm$ 0.052 for Ibc/IIb and 0.453 $\pm$ 0.047 for type II.

The differences between the two main types are well within the errors in all bands and thus statistically insignificant. The differences between Ib and Ic samples are small except in H$\alpha$, but because these samples are rather small, the errors are too large to detect a significant difference there either.

Similarly to the NCR statistics, the distribution of \emph{Fr} is flat if the CCSN distribution accurately traces the galaxy's emission, with an average value of 0.5 \citep{b3}. The average \emph{Fr}(H$\alpha$) values of all our subsamples are indeed close to this value, while in the \emph{R}-band and NUV they are slightly smaller and in the \emph{Ks}-band slightly larger. Similarly to our analysis of NCR values, we performed a K-S test to see if the radial distribution matches that of the host galaxy emission. We found that the only statistically significant difference in this analysis is between type II and a flat distribution in the \emph{Ks}-band, indicating that type II SNe preferably occur in the outer regions of galaxies, while all other distributions are formally consistent (probability over ten per cent) with a flat distribution.

Comparing types I vs. II and Ib vs. Ic in a two-sample K-S test, we found no significant difference between the \emph{Fr} distributions in any band. This result is in agreement with the comparison of the mean \emph{Fr}(H$\alpha$), \emph{Fr}(\emph{Ks}) and \emph{Fr}(NUV) values of the types. A comparison between the H$\alpha$ and NUV distributions for each type yielded no significant differences.

\subsection{Surface density profile of CCSNe}

The resulting power-law and exponential fits from our MC analysis are plotted in Fig. 4 and the parameters listed in Table 6. We obtained the following values for the parameters of these functions (for details, see Section 4.3):  $\gamma = 1.6 \pm 0.2$, $\tilde{h}_{SN} = 0.23^{+0.03}_{-0.02}$. For their sample of 224 SNe, \citet{b17} derived a normalised scale length of $\tilde{h}_{SN} = 0.29 \pm 0.01$; statistically significantly above the value we have obtained. For a Freeman disk, they obtained $\tilde{h}_{SN} = 0.32 \pm 0.03$. 

We also compared the type Ibc/IIb and type II subsamples' surface density profiles, obtaining a scale length $\tilde{h}_{SN} = 0.19^{+0.03}_{-0.02}$ for type Ibc/IIb and $\tilde{h}_{SN} = 0.29^{+0.06}_{-0.04}$ for type II, which indicates that type Ibc/IIb SNe are more centrally concentrated than type II. This indication is in agreement with previous studies: \citet{b17} obtained scale lengths of 0.24 $\pm$ 0.03 for type Ibc and 0.31 $\pm$ 0.02 for type II. In light of the different scale lengths for Hakobyan's sample and ours, and of the rapidly rising surface density of types Ibc/IIb in the central region, we obtained $\tilde{h}_{SN} = 0.24^{+0.05}_{-0.04}$ for types Ibc/IIb by ignoring the innermost radial distance bin, which agrees with \citet{b17}. The scale lengths for type II SNe in their sample and ours are consistent within the errors. The slopes of the power-law fits for the subsamples hint at type Ibc/IIb SN density falling off more rapidly as well, although the difference is not statistically significant; $\gamma$(Ibc/IIb) =  1.7 $\pm$ 0.2 and $\gamma$(II) = 1.4$^{+0.2}_{-0.3}$.

In addition, we calculated the mean normalised distances for both subsamples, presented in Table 7. For type Ibc/IIb, we obtained a mean of 0.350 $\pm$ 0.055 $R_{25}$; while for type II it is 0.476 $\pm$ 0.045. Thus, again, the type Ibc/IIb SNe seem more centrally concentrated than type II. Furthermore, we converted the corrected angular radial distances into linear distances in parsecs. Although the sizes of the galaxies vary, statistically a comparison between the linear distances for each type can be meaningful. We found a mean radial distance of 5150 $\pm$ 1220 pc for type Ibc/IIb and 6640 $\pm$ 810 pc for type II. Again there is a difference between types, but the error bars are too large to make it statistically significant. The large variations in galaxy sizes make the standard deviation of the linear distance values large as well.

Since most of the galaxies in our sample are dusty, and dust extinction affects the \emph{B}-band much more strongly than the \emph{Ks}-band, there was a concern that the \emph{B}-band $R_{25}$ might not offer a valid distance normalisation in this case. Thus, we also fitted an exponential function and a power law to distances normalised using the \emph{K}-band $R_{20}$ instead, as measured by 2MASS. While the scale lengths of these functions are not comparable due to the normalisation distances being measured at different bands -- and the scale length obtained using $R_{20}$ ($0.34^{+0.04}_{-0.03}$) is not comparable with previous results normalised using $R_{25}$ -- the power laws fitted in both cases are equal. Thus there seems to be no discrepancy.

\begin{table}
\centering
\caption{The parameters of surface density profiles of CCSNe, $\tilde{h}_{SN}$ being the scale length of an exponential function and $\gamma$ the slope of a power law on a log-log scale.}
\begin{tabular}[t]{lcc}
    \hline
         Sample & $\tilde{h}_{SN}$ & $\gamma$ \\
    \hline
    \hline
	All CCSNe ($R_{25}$) & $0.23^{+0.03}_{-0.02}$ & 1.6 $\pm$ 0.2 \\
	Hakobyan et al. 2009 & 0.29 $\pm$ 0.01 & - \\
	Freeman disk & 0.32 $\pm$ 0.03 & - \\
	Ibc/IIb & $0.19^{+0.03}_{-0.02}$ & 1.7 $\pm$ 0.2 \\
	Ibc/IIb without 1st bin & $0.24^{+0.05}_{-0.04}$ & - \\
	II & $0.29^{+0.06}_{-0.04}$ & $1.4^{+0.2}_{-0.3}$ \\
	All CCSNe ($R_{20}$) & $0.34^{+0.04}_{-0.03}$ & 1.6 $\pm$ 0.2 \\
	All CCSNe (kpc) & 3.7 $\pm$ 0.2 & $1.8^{+0.2}_{-0.1}$ \\
\hline
\end{tabular}
\end{table}

\begin{table}
\centering
\caption{The mean distances of different CCSN types from the galaxy cores, normalised to \emph{B}-band $R_{25}$ or \emph{Ks}-band $R_{20}$ and also given in parsecs. SNe in galaxies with $i \ge 70^{\circ}$ have been excluded.}
\begin{tabular}[t]{lcccc}
    \hline
         SN type & N & $R_{SN}/R_{25}$ & $R_{SN}/R_{20}$ & $R_{SN}$(pc) \\
    \hline
    \hline
	Ibc/IIb & 30 & 0.350 $\pm$ 0.055 & 0.521 $\pm$ 0.086 & 5150 $\pm$ 1220 \\
	II & 43 & 0.476 $\pm$ 0.045 & 0.695 $\pm$ 0.074 & 6640 $\pm$ 810 \\
    \hline
	Ibc & 26 & 0.367 $\pm$ 0.062 & 0.530 $\pm$ 0.094 & 5460 $\pm$ 1390 \\
	Ib & 10 & 0.385 $\pm$ 0.119 & 0.724 $\pm$ 0.221 & 7820 $\pm$ 3250 \\
	Ic & 13 & 0.379 $\pm$ 0.079 & 0.476 $\pm$ 0.101 & 4520 $\pm$ 1110 \\
    \hline
	All & 77 & 0.410 $\pm$ 0.034 & 0.608 $\pm$ 0.056 & 5840 $\pm$ 670 \\
\hline
\end{tabular}
\end{table}

\section{Discussion}

\begin{figure}
\centering
\begin{minipage}{83mm}
\includegraphics[width=8.3cm]{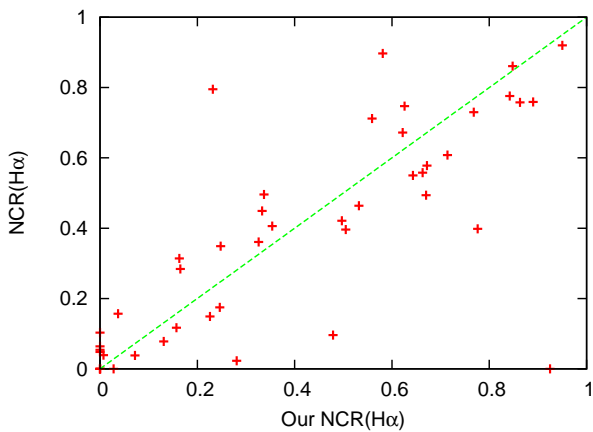}
\caption{A comparison between our NCR(H$\alpha$) values and corresponding values in A08 and A12 for those SNe also included in their study. There is no systematic difference between the two sets of values.}
\end{minipage}
\end{figure}

\subsection{Possible biases}

Comparing our NCR(H$\alpha$) values and those of A08 and A12 (this includes 46 SNe in total; see Fig. 5 for comparison), we see no systematic difference between the two sets. Most of the outlying values seem to be caused by an error in the astrometric calibration of either image. The NCR values for these SNe are 0.281 (0.023 from A08/A12) for SN 1997bs, 0.581 (0.897) for SN 1999gn, 0.925 (0) for SN 2000cr, 0.479 (0.096) for SN 2008ij and 0.776 (0.398) for SN 2009hd. We determined NCR values for the pixels surrounding the SN positions up to a radius of 1$\arcsec$, and values roughly agreeing with those in A08 and A12 were found among these pixels, suggesting a possibility of small positional errors. For SN 2000cr in particular, the NCR difference is very large, but it is located in a compact bright region. In addition, SN 2005V's value in A08 is given as 0.000. However, \citet{b3} originally gave it an NCR(H$\alpha$) of 0.861, which is consistent with our value, and we have used this for Fig. 5. The mean NCR(H$\alpha$) of these 46 overlapping SNe is 0.391 $\pm$ 0.046 from our values and 0.367 $\pm$ 0.044 from A08/A12; after removing the SNe with possible astrometric error, the mean values become 0.372 $\pm$ 0.047 and 0.383 $\pm$ 0.046 respectively. Therefore we can conclude that there is no systematic difference between the two studies.

\citet{b21} explored some possible biases that might be affecting their sample. Firstly, because of the lower peak luminosity of type II than type Ibc SNe, the former may be more difficult to detect in distant galaxies. Additionally, SNe of all types might be more difficult to detect against bright regions (especially the central regions) of host galaxies as the distance increases. The former effect could manifest itself as a decreasing ratio of type II versus type Ibc SNe with increasing host galaxy distance and the latter as a decreasing mean NCR value. \citet{b21} did detect a trend of decreasing type ratio, but explained this as simply the result of their sample selection effects: their galaxies have been chosen to yield SNe with subtype classifications, which are more unlikely to be determined for distant type II SNe because of the required systematic photometry. Furthermore, they detected no decrease in mean NCR values with increasing distance for either type Ibc or type II, and concluded that there is no selection effect favoring the detection of one over the other against bright regions and that they are thus seeing intrinsic variations in correlation.

Our sample, however, has been chosen based on \emph{galaxy} properties, paying no attention to SN types or subtypes. Therefore it is also worth checking whether we see any trends with increasing distance. The mean NCR values from H$\alpha$ and \emph{Ks}-band images are plotted in Fig. 6 as a function of host galaxy distance. Like \citet{b21}, we detect no statistically significant decrease in NCR values with distance in either subsample or the full sample in H$\alpha$ or the \emph{Ks}-band. The values show some variation, but no obvious trends. Therefore we agree with \citet{b21} that there seems to be no selection effect favoring the detection of type Ibc/IIb SNe over type II \emph{against bright regions} within our distance limit. In addition, the ratio of type II SNe versus Ibc/IIb does not significantly drop with increased distance: splitting the sample in half based on distance, we find a ratio of 1.6 $\pm$ 0.5 for nearby and 1.3 $\pm$ 0.4 for distant galaxies. The Poissonian error bars of the ratios are quite large, and with a sample of this size this result is only tentative.

\begin{figure*}
\centering
\begin{minipage}{87mm}
\includegraphics[width=8.7cm]{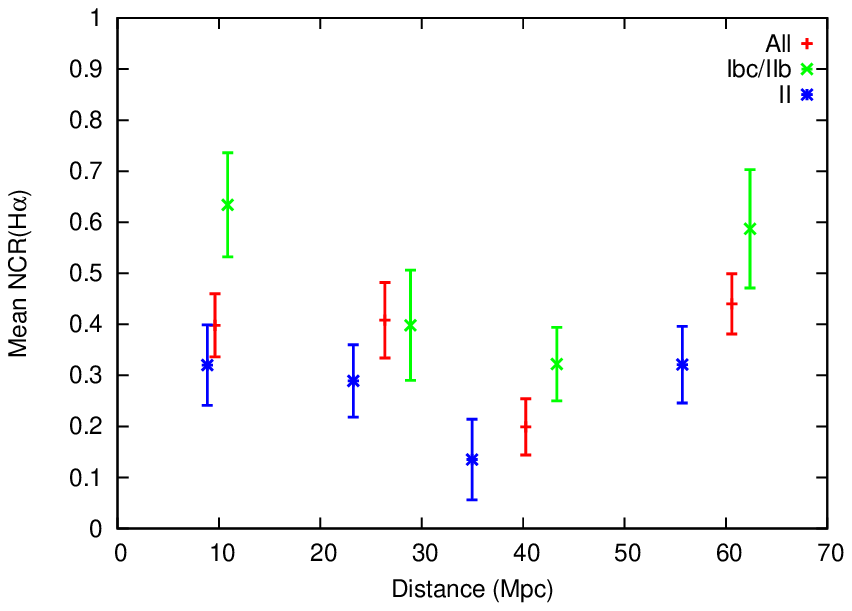}
\end{minipage}
\begin{minipage}{87mm}
\includegraphics[width=8.7cm]{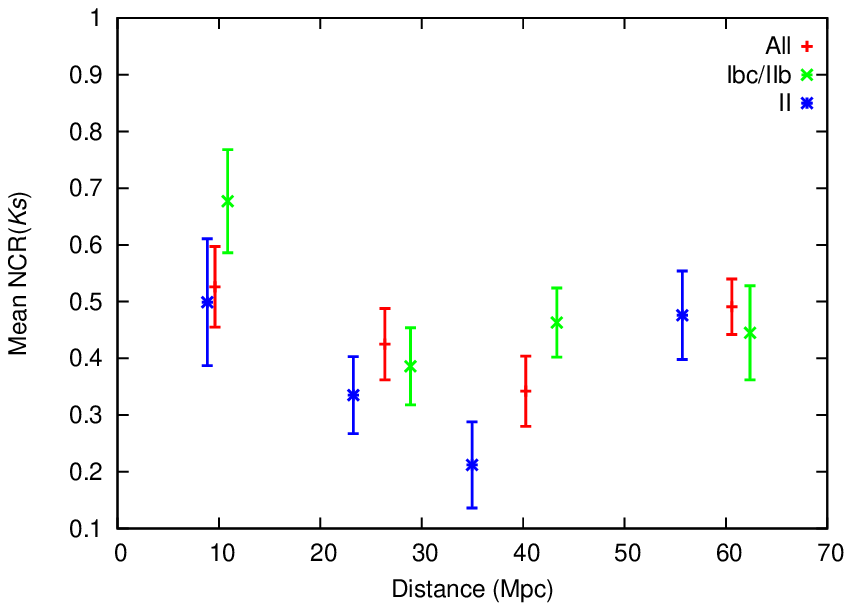}
\end{minipage}
\caption{The mean NCR values from H$\alpha$ (left) and \emph{Ks}-band (right) images.}
\end{figure*}

Furthermore, we have compared the ratio of the main types in our sample to the volumetric SN rates inside a recession velocity limit of 2000 km s$^{-1}$ from the Lick Observatory Supernova Search (LOSS) \citep{b49}, and to the relative numbers of SNe within the same limit, discovered betwen 1998 and 2012.25 and compiled by \citet{b43}. The ratio of type II vs. types Ibc/IIb in our sample is $1.5 \pm 0.3$, while from the volume-limited LOSS rates we find a ratio of 1.7 $\pm$ 0.4 and from \citet{b43} 1.6 $\pm$ 0.3. Based on this, we do not seem to have missed a significant number of type II SNe \emph{compared to types Ibc/IIb}.

There are significant amounts of dust in the centres of IR-bright galaxies, which can cause up to 80 per cent of SNe to be missed in the nuclear regions of LIRGs (e.g. Mattila et al. 2012; Miluzio et al. 2013). As can be inferred from Fig. 3, the \emph{Ks}-band \emph{Fr} values show a general central deficit compared to the optical bands. This deficit is dominated by merger and galaxy pair systems, where our sample only includes one SN with \emph{Fr}(\emph{Ks}) $<$ 0.23 (19 SNe in total in these galaxies). The indication is that most of the central SNe are missed due to extinction effects for the optical SN searches, but the central \emph{Ks}-band light is not due to the reduced extinction in this band. By eye, the CDF is smoothed out by adding roughly twelve SNe in the \emph{Fr}(\emph{Ks}) range 0 -- 0.25 in the merger sample, resulting in a missed central SN fraction of roughly 90 per cent, which is consistent with the estimates presented by \citet{b10}. In the \emph{R}-band there is no such deficit, because the central emission suffers from similar extinction as the central SNe. If all the SNe were to be detected, the CDF in \emph{Ks}-band would resemble the one in \emph{R}-band -- now we see the \emph{Ks}-band emission but have not detected the SNe. This effect concerns all types of SNe equally, and thus any bias to our analysis should be insignificant.

\subsection{Differences between the SN types}

Type Ibc/IIb SNe occur significantly closer to regions of recent star formation than type II SNe. The average NCR(H$\alpha$) value for Ibc/IIb is 0.475 $\pm$ 0.052, which indicates they closely follow the distribution of H$\alpha$ light. Comparing the distribution of type Ibc/IIb SNe and a flat distribution of NCR values with the K-S test confirms this association. Type II SNe, on the other hand, do not follow the star-forming regions. A K-S test between the two main types indicates that they arise from different stellar populations. Earlier studies (e.g. A08, A12) explain this with type Ibc SNe having more massive progenitors which thus have shorter lifetimes and therefore have less time to drift away from their birthplaces, or the H II region has less time to dissipate, leading to a closer association with star-forming regions. In addition, as more SNe occur in an H II region, it will gradually cease to exist due to loss of stars emitting ionizing radiation and strong stellar winds and the SN explosions blowing away the gas. Therefore the first SNe would show a stronger association with the region (as discussed by A12).

When we examine the subtypes Ib and Ic separately, we find that the NCR(H$\alpha$) values of our type Ib SNe are consistent with those of type II, while those of type Ic are confirmed to be significantly higher -- too high, in fact, to be randomly drawn from the H$\alpha$ emission; they are strongly correlated with the H II regions. The difference between types Ib and II is rather small in A12 as well -- in fact, the error limits of their mean NCRs overlap slightly. This indicates an overlapping mass range for the progenitors of types Ib and II, or at least that they both are outside the lifetime range traced by H$\alpha$ emission.

A caveat was recently discussed by \citet{b38}. He argued that since individual small H II regions would dissipate in a time scale shorter than the ages of the progenitors of type Ibc/IIb or II SNe, no meaningful constraints could be obtained for stars less massive than 50 -- 100 $M_{\odot}$; in addition, small H II regions would not be detected in the ground-based images used for this study. On the other hand, giant H II regions would last on the order of 20 Myr, meaning that one can only make weak constraints based on association with H$\alpha$ emission; namely, limits of $\geq 12 M_{\odot}$ for SE-SN progenitors and $\geq 8 M_{\odot}$ for type II, based on what is considered the smallest possible mass for a SN progenitor. The differences in correlation with H$\alpha$ light that are seen in both our study and those in \citet{b3} and later papers do indicate that type Ic progenitors are more massive than the $12 M_{\odot}$ limit while some Ib and II progenitors could be less massive.

One way to address this problem is to also use UV light as an indicator of \emph{recent} -- on timescales of 16 to 100 Myr \citep{b42} -- as opposed to \emph{on-going}, star formation. This was done for type II-P and IIn SNe and `impostors' in A12. Based on our results, we find that type II SNe accurately trace the galaxy NUV light with a mean NCR(NUV) close to 0.5. We obtained a mean NCR(NUV) close to 0.5 (consistent with type II) for type Ib and a significantly higher one (confirmed by the K-S test against a flat distribution) for Ic. Thus type Ic SNe preferably occur in regions with high NUV intensity. This lends additional support to type Ic progenitors being of higher mass than those of types Ib and II. Since the difference between types Ib and II in both H$\alpha$ and NUV is insignificant and they both are consistent with being randomly drawn from the NUV light distribution, this indicates that their progenitor lifetimes are $>$ 16 Myr as traced by the NUV \citep{b42}. Lower-mass interacting binaries, instead of massive Wolf-Rayet stars, have been proposed to be the progenitors of a significant fraction of type Ibc SNe (e.g. Podsiadlowski, Joss \& Hsu 1992; Smartt 2009) and of type IIb SNe \citep{a12} -- which is also supported by a lack of direct detections of their progenitors \citep{b43}. \citet{b49} and \citet{a02} suggested that type Ib and IIb SNe arise mostly from interacting binaries and type Ic SNe at least partly from more massive single WR stars. This could explain our results.

Our \emph{R}-band results are very similar to NUV. The K-S test shows a significant difference between the SNe of types Ibc/IIb -- and specifically Ic -- and a flat distribution of NCR(\emph{R}) values. The calculated mean values for types II and Ib are close to 0.5, while Ic SNe are concentrated into the bright areas. It is likely, however, that \emph{R}-band light does not provide a useful indicator on SN environments, since it includes both the H$\alpha$ line emission and a stellar continuum from different sources.

\emph{Ks}-band light is thought to be a good tracer of stellar mass i.e. old stars \citep{b9} since, in normal galaxies, most of the mass is in older stars. However, the contribution of red supergiants to the \emph{Ks}-band light, which also exists in normal spiral galaxies \citep{b37}, dominates this band in starburst galaxies \citep{b39}, which would make \emph{Ks}-band light a useful indicator of recent star formation similar to H$\alpha$. In the \emph{Ks}-band our results are indeed similar to those in H$\alpha$: the mean NCR for type Ic is significantly larger than for type II, while types Ib and II are consistent with each other. The K-S test confirms this. Additionally there is a significant difference from a flat distribution for type II, but not for other types. Furthermore, \emph{Ks}-band emission is much less affected by extinction than H$\alpha$, which we can see in the difference between \emph{Fr} values; smaller \emph{Fr} values could be expected in H$\alpha$ than in \emph{Ks}-band because of the central H$\alpha$ flux being strongly absorbed by the dust. Extinction could have an effect on NCR values as well. However, this similarity in NCR values suggests that the lack of extinction correction makes no great difference here, assuming that \emph{Ks}-band light does trace the birthplaces of massive stars in actively star-forming galaxies. 

As for the radial (\emph{Fr}) distributions of the CCSNe, the K-S test finds no statistically significant excess of type Ibc/IIb SNe in the inner regions of galaxies compared to the host galaxy emission in the optical bands, which can also be seen in Fig. 3. Our mean \emph{Fr}(H$\alpha$) values are consistent with those of the SNe in \citet{b31} within the errors for all types, however. In the \emph{Ks}-band we see a difference between type II and a flat distribution, but no difference between types Ibc/IIb and a flat distribution. However, the radial distribution of optically-detected SNe is not ``supposed'' to be flat in this band (see Sect. 6.1), and thus the difference between type II SNe and a flat distribution does not necessarily indicate an outer excess of type II SNe. 

However, the mean normalised radial distances and the scale length of the SN surface density are significantly shorter for types Ibc/IIb than for type II. These results indicate that SNe of type Ibc/IIb are indeed more concentrated to the central regions of galaxies than those of type II, which is in agreement with previous results by e.g. \citet{b17} and \citet{b31}. Furthermore, the radial distances for types Ib and Ic are equal within the errors, meaning that here we do see a difference between types Ib and II. The different results from these two tests may be explained by the fact that they measure different things. The normalised radial distances are not affected by the properties of the central regions, while \emph{Fr} values are.

\subsection{Differences between normal spirals and IR-bright galaxies}

The clearest difference we see in the distributions of CCSNe in normal vs. IR-bright galaxies is in the scale length of their surface density. Based on a Monte Carlo analysis, we find a normalised scale length of $\tilde{h}_{SN} = 0.23^{+0.03}_{-0.02}$, which is statistically significantly smaller than that derived by \citet{b17} for a sample of CCSNe in spiral galaxies (0.29 $\pm$ 0.01). However, there are roughly five ``extra'' Ibc SNe in the central regions compared to what one would expect from an exponential surface density profile derived by ignoring the innermost radial distance bin. Since by ignoring the central bin we obtained a similar scale length ($0.24^{+0.05}_{-0.04}$) for types Ibc/IIb as \citet{b17} did for their type Ibc sample ($0.24 \pm 0.02$), and since the type II scale lengths match within the errors (we get $\tilde{h}_{SN} = 0.29^{+0.06}_{-0.04}$ and \citet{b17} report 0.31 $\pm$ 0.02), the indication is that our shorter overall scale length is mainly due to an excess of types Ibc/IIb in the central regions, with respect not only to type II but also to types Ibc/IIb in normal galaxies. 

Prior radial analysis papers, and our corresponding results, indicate more central positions for SNe of type Ibc/IIb than for other type II. A metallicity gradient is often brought up as an explanation (e.g. Anderson \& James 2009; Hakobyan et al. 2009), since metallicity is known to be a major factor in the mass-loss rates of the progenitor stars (e.g. Crowther et al. 2002).

``Disturbed'' galaxies, especially merger systems, have a less pronounced metallicity gradient than normal isolated galaxies (e.g. Rich et al. 2012). Dividing the galaxy sample into isolated, disturbed and merging/closely-paired systems (see Sect. 2), we see that type Ibc SNe are more centrally located in ``disturbed'' galaxies. The mean $R_{SN}/R_{25}$ for types Ibc/IIb is 0.588 $\pm$ 0.138 in isolated galaxies, 0.315 $\pm$ 0.085 in ``disturbed'' ones and 0.227 $\pm$ 0.050 in merging/close-pair systems. Thus type Ibc/IIb SNe in mergers and ``disturbed'' galaxies are significantly more centralized. This is consistent with what is seen by \citet{b31}, but it is the opposite of what one would expect if metallicity was the deciding factor, since the central metallicity in ``disturbed'' galaxies is expected to be \emph{lower} than in isolated galaxies. \citet{a04} also show that the metallicity gradient gets less pronounced with increased IR luminosity. However, we divided our sample into (relatively) IR-luminous and IR-faint galaxies, the threshold being the median log $L_{FIR}$ of our sample (10.57) with 44 SNe in the brighter half and 42 in the fainter half. We find that the radial distances for type Ibc/IIb SNe are the same, within the errors, regardless of which subsample the host galaxy belongs to (mean $R_{SN}/R_{25}$ = 0.359 $\pm$ 0.089 in the fainter galaxies and 0.344 $\pm$ 0.072 in the brighter ones). Thus we see no evidence of a metallicity effect driving the relative centralization of type Ibc/IIb SNe.

In the study of \citet{b31} the centralization is taken to indicate that in the central regions of ``disturbed'' galaxies, the IMF is top-heavy. Klessen, Spaans \& Jappsen (2007) presented a model, based on numerical simulations, where the increased temperature in intensely star-forming regions causes the local Jeans mass to increase, which would in turn result in a top-heavy IMF with a peak around 15 $M_{\odot}$. Obviously this would affect strongly star-forming galaxies more than normal spirals. However, we are seeing both Ib and Ic SNe in the central regions (four of type Ib and five of type Ic inside 0.2 $R_{25}$; see also Table 7). According to our NCR results, the mass ranges of the progenitors of type II and Ib SNe seem to overlap strongly. Thus, while the enhanced fraction of type Ic SNe can be explained by a modified IMF in the central regions, that of type Ib SNe seemingly cannot. Thus our results suggest that while the top-heavy IMF may play a role, it is not the only factor behind the enhanced SE fraction in the central regions.

Since we see no evidence of a metallicity dependence in the centralization, and since relatively low-mass binaries probably form a significant fraction of type Ib(c) progenitors (see Section 6.2), we suggest that an enhanced close binary fraction in areas of high star formation density could be a significant factor, perhaps along with the top-heavy IMF to account for the prevalence of both Ib and Ic SNe. \citet{a06} showed that in young star-forming regions in the Milky Way, the fraction of binary systems as a whole is either very weakly decreased or not at all compared to the field, while the fraction of \emph{close} binaries is significantly enhanced. Furthermore, a large fraction of star formation happens in young massive star clusters \citep{a08}, which have been shown by \citet{a09} to correlate with X-ray binaries. Our sample galaxies tend to have strong central star formation, meaning that this would also apply to the central regions. In such dense regions found inside clusters, primordial hard binaries (those with high binding energy) tend to harden and soft ones tend to soften according to Heggie's law \citep{a07}. While some of these binaries do get disrupted, this probably will not happen until many of them have exploded as SNe.

The relative numbers of SNe of different types might also offer an indicator of differences between IR-bright and normal galaxies. As already stated in Section 6.1, the ratio of the numbers in our main subsamples (II vs. Ibc/IIb) should be $1.6 \pm 0.3$ (or 1.7 $\pm$ 0.4), based on the volume-limited numbers of different types of SNe inside the recession velocity limit of 2000 km s$^{-1}$, according to \citet{b43} (or \citet{b49}). In our sample the observed ratio is $1.5 \pm 0.3$, and thus in agreement with the volume-limited samples. If the central ``extra SNe'' mentioned above are excluded, our type ratio would be $1.6 \pm 0.4$. This indicates the rate of type Ibc/IIb SNe compared to type II in IR-bright galaxies is not significantly different than in normal galaxies. The suggested excess of type Ibc/IIb SNe in the central regions \emph{compared to normal galaxies} is thus not strong enough to affect the relative rates significantly.

\section{Conclusions}

We have presented results for correlations between CCSNe of different types and light in the H$\alpha$, \emph{R}-band, \emph{Ks}-band and NUV in IR-bright galaxies. We have also presented an analysis of radial positions of these SNe in relation to the host galaxy light, and fitted functions to describe the SN surface densities. We have, when possible, compared these results with earlier studies, with samples dominated by normal spiral galaxies.

Our results confirm that type Ic SNe show a higher correlation with H$\alpha$ light (and thus on-going star formation) than types II or Ib. In the NUV (recent star formation) types II and Ib accurately trace the host galaxy light, while Ic SNe are concentrated in the bright regions. Assuming that a higher correlation with H$\alpha$ light is because of a shorter lifetime of the progenitor star and thus a higher mass, the average masses of type Ic progenitors are indicated to be higher than those of types Ib and II, as asserted previously by e.g. A08 and A12. However, both type Ib and II are consistent with being randomly drawn from the NUV light distribution but not from the H$\alpha$ distribution; thus the mass difference between these progenitors may not be significant, and instead binarity may play a larger role.

In the \emph{Ks}-band the pixel statistics results are quite similar; type Ibc/IIb SNe are more strongly correlated with the underlying galaxy light. Since the \emph{Ks}-band light of actively star-forming galaxies is dominated by red supergiants, it is also a tracer of recent star formation in these galaxies. Therefore we can infer the same results as from the H$\alpha$ and NUV data, further supported by the weaker extinction in the infrared.

Type Ibc/IIb SNe are preferentially more centrally located in their host galaxies than type II in our sample, which is consistent with previous results by \citet{b31}. This is probably due to the effect of an enhanced close binary fraction and/or a top-heavy IMF in the central regions compared to the outer regions (however, the IMF \emph{alone} seems inadequate to explain our results). We see no evidence of metallicity playing a major role here. A significant fraction of SNe may be missed in the centres of IR-bright galaxies due to high line-of-sight extinctions and contrast effects, which we see clearly when comparing the CCSN distribution with that of host galaxy \emph{Ks}-band light, but we do not find any evidence that type II SNe would be more strongly affected by this than type Ibc/IIb in our sample.

The scale length of the CCSN surface density is significantly smaller in IR-bright galaxies than in samples dominated by normal spiral galaxies. Furthermore, this seems to be caused mainly by an excess of type Ibc/IIb SNe in the central regions, compared not only to type II but to normal spiral galaxies as well: by ignoring the central SNe the scale lengths for both main types (Ibc/IIb and II) match those derived by \citet{b17}. The top-heavy IMF and/or enhanced close binary fraction thus especially apply to IR-bright galaxies, many of which are ``disturbed''. If dust extinction in these galaxies is greater than normal and more (central) SNe are missed, this effect is even stronger than what is seen here.

\section*{Acknowledgements}

We thank the anonymous referee for his or her very useful comments. We also thank Phil James and Joe Anderson for their comments and constructive discussion, Stacey Habergham for sharing the \emph{Fr} values from her paper, Rubina Kotak and Justyn Maund for discussion on biases, Andrew Thompson for providing a copy of his MSc thesis for that discussion, and Jos\'e Ramon S\'anchez-Gallego and Johan Knapen for sharing their optical images of four galaxies. EK acknowledges financial support from the Jenny and Anti Wihuri Foundation.

Based on observations made with the Nordic Optical Telescope, operated on the island of La Palma jointly by Denmark, Finland, Iceland, Norway, and Sweden, in the Spanish Observatorio del Roque de los Muchachos of the Instituto de Astrofisica de Canarias. The data presented here were obtained in part with ALFOSC, which is provided by the Instituto de Astrofisica de Andalucia (IAA) under a joint agreement with the University of Copenhagen and NOTSA. The William Herschel Telescope is operated on the island of La Palma by the Isaac Newton Group in the Spanish Observatorio del Roque de los Muchachos of the Instituto de Astrof\'isica de Canarias. 

This research has made use of the NASA/IPAC Extragalactic Database (NED) which is operated by the Jet Propulsion Laboratory, California Institute of Technology, under contract with the National Aeronautics and Space Administration. We acknowledge the usage of the HyperLeda database (http://leda.univ-lyon1.fr).

\label{lastpage}

\end{document}